\documentclass[useAMS,usenatbib,aastex]{mn2epreprint}
\usepackage[a4paper,centering,totalwidth=520pt,totalheight=683pt]{geometry}
\usepackage{graphics}
\usepackage{graphicx}
\usepackage{amssymb}
\usepackage{natbib}
\usepackage[bookmarks=true, bookmarksnumbered=true, colorlinks=true,
urlcolor=darkblue, citecolor=darkpowderblue, linkcolor=darkred,
breaklinks=true]{hyperref} 

\providecommand{\adsurl}[1]{\href{#1}{ADS}}

\usepackage{bm}
\usepackage{dcolumn}
\usepackage{color}

\def\beq{ \begin{equation}}
\def\eeq{\end{equation} }
\def\bea{\begin{eqnarray}}
\def\eea{\end{eqnarray}}

\definecolor{darkblue}{rgb}{0,0,0.5}
\definecolor{darkpowderblue}{rgb}{0,0.2,0.6}
\definecolor{darkred}{rgb}{0.55,0,0}

\begin{document}

\title[Subhalo properties and DM annihilation]{Characterization of subhalo structural properties and implications for dark matter annihilation signals}
\author[A. Molin\'e et al.]
{\'Angeles Molin\'e$^{1, 3}$\thanks{maria.moline@ist.utl.pt}, 
Miguel A. S\'anchez-Conde$^{2}$\thanks{sanchezconde@fysik.su.se}, 
Sergio Palomares-Ruiz$^{3}$\thanks{sergiopr@ific.uv.es}, 
\newauthor
Francisco Prada$^{4,5,6}$\thanks{f.prada@csic.es}\\
$^{1}$ CFTP, Instituto Superior T\'ecnico, Universidade T\'ecnica de Lisboa,  
	Av. Rovisco Pais 1, 1049-001 Lisboa, Portugal \\
$^{2}$ Oskar Klein Centre for Cosmoparticle Physics, Department of Physics, Stockholm University, SE-10691 Stockholm, Sweden \\
$^{3}$ Instituto de F\'{\i}sica Corpuscular (IFIC),  CSIC-Universitat de Val\`encia,  Apartado de Correos 22085, E-46071 Valencia, Spain\\
$^{4}$ Campus of International Excellence UAM+CSIC, Cantoblanco, E-28049 Madrid, Spain\\
$^{5}$ Instituto de F\'{\i}sica Te\'orica (UAM/CSIC), Universidad Aut\'onoma de Madrid, Cantoblanco, E-28049 Madrid, Spain\\
$^{6}$ Instituto de Astrof\'{\i}sica de Andaluc\'{\i}a (IAA-CSIC), Glorieta de la Astronom\'{\i}a, E-18008 Granada, Spain}
\preprint{CFTP/16-006, IFIC/16-15}

\date{}

\pubyear{2017}

\maketitle

\label{firstpage}

\begin{abstract}
A prediction of the standard $\Lambda$CDM cosmology is that dark matter (DM) halos are teeming with numerous self-bound substructure, or subhalos. The precise properties of these subhalos represent important probes of the underlying cosmological model. We use data from \emph{Via Lactea II} and \emph{ELVIS}  N-body simulations to learn about the structure of subhalos with masses $10^6 - 10^{11}~h^{-1} M_{\odot}$. Thanks to a superb subhalo statistics, we study subhalo properties as a function of distance to host halo center and subhalo mass, and provide a set of fits that accurately describe the subhalo structure. We also investigate the role of subhalos on the search for DM annihilation. Previous work has shown that subhalos are expected to boost the DM signal of their host halos significantly. Yet, these works traditionally assumed that subhalos exhibit similar structural properties than those of field halos, while it is known that subhalos are more concentrated. Building upon our N-body data analysis, we refine the substructure boost model of \citet{Sanchez-Conde:2013yxa}, and find boosts that are a factor 2-3 higher. We further refine the model to include unavoidable tidal stripping effects on the subhalo population. For \emph{field} halos, this introduces a moderate ($\sim20\%-30$\%) suppression. Yet, for subhalos like those hosting dwarf galaxy satellites, tidal stripping plays a critical role, the boost being at the level of a few tens of percent at most. We provide a parametrization of the boost for field halos that can be safely applied over a wide halo mass range.
\end{abstract}


\section{Introduction}
\label{sec:intro}

Small matter density perturbations in the early Universe grow via gravitational instability giving rise to cold dark matter (CDM) bound structures known as \emph{halos}. In the standard $\Lambda$CDM cosmological framework \citep{Ade:2015xua}, structure formation proceeds hierarchically (see, e.g., \citet{Frenk:2012ph}), with low-mass halos forming first and then large-mass halos resulting from the merging and accretion of those smaller halos. Yet, not all of the smaller structures are destroyed in this process and, indeed, the hierarchical nature of CDM structure formation implies that halos are teeming with self-bound substructure, namely \emph{subhalos}, which orbit within the potential well of a more massive \emph{host} halo.

The study of the statistical and structural properties of the subhalo population is of prime importance because subhalos represent important probes of the mass accretion history and dynamics of host halos and thus, ultimately, of the underlaying cosmological model. For instance, the comparison between subhalo results derived from N-body simulations and observational data has revealed potential issues in our understanding of galaxy formation, e.g., the missing satellites problem \citep{Klypin:1999uc, Moore:1999nt} and the too-big-to-fail problem \citep{BoylanKolchin:2011de, BoylanKolchin:2011dk, Tollerud:2014zha, Garrison-Kimmel:2014vqa}. Subhalos may also play a crucial role for indirect dark matter (DM) searches, namely, the detection of the annihilation products of DM particles. One of the most plausible DM candidates is a weakly interacting massive particle (WIMP), with mass lying from the GeV to the TeV scale (see, e.g., \citet{Bertone:2004pz}). WIMPs are assumed to be stable in cosmological time scales and to annihilate into Standard Model particles. Once small density perturbations in the early Universe grow via gravitational instability and halos are formed, the density of WIMPs is orders of magnitude larger than the average matter density and the annihilations of WIMPs occurring in halos and subhalos, generate gamma-rays, antimatter and neutrinos. In these searches, the signal is proportional to the square of the DM density and hence, the presence of substructure could produce an enhancement (or \emph{boost}) over the expected signal from the smooth distribution of DM in the host halo. Moreover, the closest of these subhalos might represent by themselves prime targets for indirect DM detection \citep{Baltz:1999ra, Tasitsiomi:2002vh, Koushiappas:2003bn, Diemand:2005vz, Baltz:2006sv, Pieri:2007ir, Kuhlen:2008aw, Buckley:2010vg, Belikov:2011pu, Ackermann:2012nb, Mirabal:2012em, Zechlin:2012by, Berlin:2013dva, Bertoni:2015mla, Schoonenberg:2016aml}.

Different techniques have been proposed and used to learn about the properties of subhalos. Observationally, most of them are limited to our Local Group. The most massive subhalos host low-luminosity dwarf galaxies, so by studying the velocity dispersion profiles of their stars, the underlying DM distribution and thus, the subhalo internal structure, could be inferred \citep{Lokas:2004sw, Strigari:2006rd, Martinez:2009jh}. Moreover, by quantifying the number, velocities and distances to the host halo centers of these dwarf satellite galaxies, it is possible to infer statistical properties of the whole subhalo population \citep{Prada:2003ep, BoylanKolchin:2011de}. Another way to study smaller or more distant subhalos is by means of gravitational lensing. The alteration of the expected surface brightness distribution of a lensed image when a subhalo-size structure is close to this arc, could help us to shed light on the subhalo mass distribution \citep{Suyu:2010kd}.\footnote{Although due to degeneracies, measuring its density profile by only using this technique is not possible \citep{Schneider:2013sxa}.} On the other hand, a complementary effect caused by DM substructure is the production of time delay perturbations in gravitationally lensed systems, which are free of the referred degeneracies \citep{Keeton:2008gq}. 

Traditionally, N-body cosmological simulations have also been used to study DM halo substructure. Although the study of subhalo properties represents a computational challenge due to the very large dynamic range which is required, during the last years improved N-body cosmological simulations have proven to be crucial for understanding the properties of substructure within Milky Way-size systems in a $\Lambda$CDM Universe, as is the case of the \emph{Aquarius Project} \citep{Springel:2008cc}, \emph{Via Lactea} \citep{Diemand:2006ik},  \emph{Via Lactea II} (VL-II) \citep{Diemand:2008in}, \emph{Exploring the Local Volume in Simulations} (ELVIS) \citep{Garrison-Kimmel:2013eoa} and GHALO \citep{Stadel:2008pn}. These simulations have revealed in great detail the existence of a population of subhalos in larger halos, even at high redshifts. Simulations of larger volumes have also allowed us to study substructure properties at different (higher) mass scales, as Bolshoi \citep{Klypin:2010qw}, BolshoiP and MultiDark (\citet{Prada:2011jf}, P12; \citet{Riebe:2011gp, Klypin:2014kpa, Rodriguez-Puebla:2016ofw}) and, more recently, \emph{Copernicus Complexio} (COCO) \citep{Hellwing:2015upa}. Yet, as expected, all mentioned simulations are limited by the particle mass resolution and, indeed, are far from resolving the whole subhalo hierarchy by orders of magnitude in mass. On the other hand, semi-analytical studies, based on the extended Press-Schechter formalism \citep{Press:1973iz, Bond:1990iw, Sheth:1999su}, do not share these problems, although the relevant gravitational physics can only be treated in an approximate way \citep{Taylor:2000zs, Benson:2001at, Ullio:2002pj, Taffoni:2003sr, Zentner:2004dq, Penarrubia:2004et, Oguri:2004mw, Taylor:2004gq, Taylor:2004gp, Taylor:2004gq, vandenBosch:2004zs, Zentner:2005jr, Kuhlen:2008aw, Gan:2010ax, Pieri:2009je, Purcell:2012kd}.

During the last two decades, N-body cosmological simulations have shown that it is possible to describe the DM density distribution of halos with a single universal parametrization \citep{Navarro:1995iw, Navarro:1996gj, Moore:1999gc, Klypin:2000hk, Navarro:2008kc}.\footnote{However, see, e.g., \citet{Graham:2005xx, Hjorth:2015bfa} for a discussion on the non-universality of DM density profiles.} The DM density profiles can be fully determined with a set of critical parameters. For NFW profiles \citep{Navarro:1995iw, Navarro:1996gj}, for instance, the halo mass and the so-called halo \emph{concentration} fully determines the halo internal structure. In practice, NFW fits to the DM distribution inside halos as measured in N-body simulations allow us to infer the value of the halo concentration. The concentration of field halos has been extensively studied in the past and several concentration-mass relations, $c(M)$, have been proposed in the literature (see, e.g., \citet{Bullock:1999he, Hennawi:2005bm, Neto:2007vq, Duffy:2008pz, Maccio':2008xb, MunozCuartas:2010ig, Prada:2011jf, Ludlow:2013vxa, Diemer:2014gba, Correa:2015dva}). In contrast, subhalo concentrations remain more uncertain. For instance, no functional form has been proposed for the subhalo concentration-mass relation, $c_{\rm sub}(m_{\rm sub})$, up to now. Some of the reasons have to do with the difficulty in defining and assigning concentrations to subhalos in simulations. As a result, for computing the substructure boost to the DM annihilation signal, a common practice in the past has been the use of the concentration derived from {\it field} halos as the concentration of subhalos of the same mass (see, e.g., \citet{Lavalle:1900wn, Kuhlen:2008aw, Charbonnier:2011ft, Pinzke:2011ek, Gao:2011rf, Nezri:2012tu, Anderhalden:2013wd, Sanchez-Conde:2013yxa, Ishiyama:2014uoa}). Although this assumption represents a reasonable first order approximation, the current status of the field is calling for a more refined substructure boost model that relies on more accurate subhalo concentration values. Indeed, N-body simulations have unequivocally shown that subhalos exhibit higher inner DM densities and are on average more concentrated than field halos of the same mass (see, e.g., \citet{Ghigna:1999sn, Bullock:1999he, Ullio:2002pj, Diemand:2007qr, Diemand:2008in, Diemand:2009bm}).

In this work, we address some of these questions in detail by making use of public data from the VL-II and ELVIS N-body cosmological simulations. Altogether, these simulations allow us to study the subhalo internal properties over several orders of magnitude in subhalo mass. In addition, thanks to their superb halo statistics, they make possible a careful study of subhalo properties as a function of the distance to the host halo center, $r$. As a result, we are able to propose an accurate fit for $c_{\rm sub}(m_{\rm sub}, r)$, the first one of its kind to our knowledge. We will then use the $c_{\rm sub}(m_{\rm sub}, r)$ relation derived from the results of the VL-II and ELVIS simulations to compute and update the substructure boost to the total annihilation signal.

The work is organized as follows. In section \ref{sec:Nbody} we start by defining the most useful halo and subhalo quantities and by briefly describing the N-body cosmological simulation data sets that we use, i.e., VL-II and ELVIS. Later, in the same section, we present in detail our analysis of subhalo concentrations and provide best fits as a function of radial distance to the host halo center and of subhalo mass. We also quantify the associated subhalo-to-subhalo scatter found in the simulations. Section \ref{sec:boost} is devoted to the calculation of the boost to the DM annihilation signal due to subhalos, by means of the results found in section \ref{sec:Nbody}. This new substructure model should be perceived as a refinement of the one in \citet{Sanchez-Conde:2013yxa}. We also provide accurate fits to the boost. We conclude in section \ref{sec:summary} with a summary of our main results.


\section{Inferring subhalo properties from N-body cosmological simulations} 
\label{sec:Nbody}

\subsection{Definition of halo and subhalo properties}  
\label{sec:definitions}

A more formal definition of the halo concentration is $c_{\Delta} \equiv R_{\rm{vir}}/r_{-2}$, i.e., the ratio of the halo virial radius, $R_{\rm{vir}}$, and the radius $r_{-2}$ at which the logarithmic slope of the DM density profile $\frac{d \log\rho}{d \log\,r}=-2$. The virial radius at redshift $z$ is defined as the radius that encloses a halo mean density $\Delta$ times the critical (or mean, depending on the chosen convention) density of the Universe, $\rho_c(z)$. This standard definition of halo concentration, while very useful for the study of the internal structure of well-resolved halos, is directly less suitable for subhalos for several reasons. On one hand, the virial radius of subhalos is not well defined. Tidal stripping removes mass from the outer parts of subhalos and, as a result, subhalos are truncated at smaller radii compared to field halos of the same mass \citep{Ghigna:1998vn, Taylor:2000zs, Kravtsov:2004cm, Diemand:2006ik, Diemand:2007qr}. The subhalo DM density profiles thus drop very steeply near the edge of the subhalo (see, e.g., \citet{Kazantzidis:2003hb}). On the other hand, although the central parts of the subhalo are expected to be unaffected by mass loss \citep{Diemand:2008in}, the particle resolution of current simulations does not allow for an accurate description of subhalo density profiles in the innermost regions of the subhalos and of the host halo (see, e.g., the discussion in \citet{Diemand:2009bm}). Therefore, describing the structural properties of a subhalo is not a trivial task and it becomes highly desirable to find a definition for the subhalo concentration which is independent of any density profile and of the particular definition used for the virial radius.

One such way to characterize the concentration parameter is to express the mean physical density, $\bar{\rho}$, within the radius of the peak circular velocity $V_{\rm{max}}$, in units of the critical density of the Universe at present, $\rho_{c}$, as \citep{Diemand:2007qr, Diemand:2008in, Springel:2008cc}
\begin{equation}
c_{\rm V}=\frac{\bar{\rho}(R_{\rm{max}})}{\rho_{c}} = 2\left( \frac{V_{\rm{max}}}{H_{0} \, R_{\rm{max}}}\right)^{2}\,,
\label{eq:cvrho}
\end{equation}
where $R_{\rm{max}}$ is the radius at which $V_{\rm{max}}$ is attained and $H_0$ is the Hubble constant. Note that, in this way, $c_{\rm V}$ can be directly obtained independently of the assumed form for the subhalo DM density profile. At the same time, $c_{\rm V}$ still fully encodes the essential meaning attached to the traditional concentration parameter. Moreover, $V_{\rm{max}}$ is less affected by tidal forces \citep{Kravtsov:2004cm, Diemand:2007qr}. 

Yet, finding a relation between $c_{\Delta}$ and $c_{\rm V}$ is convenient in order to facilitate both a better intuition on subhalo concentration values and to compute annihilation boost factors in Sec.~\ref{sec:boost}, and ultimately, for a better comparison with previous works. This $c_{\Delta} - c_{\rm V}$ relation, though, will necessarily rely on the assumption of a particular functional form for the DM density profile. 

For spherical (untruncated) subhalos, the virial mass, $m_{\Delta}$, at redshift $z=0$, is defined as
\begin{equation}
\label{eq:D}
m_{\Delta}  = \frac{4\pi}{3} \, r_{\Delta}^{3} \, \rho_c \, \Delta ~,
\end{equation}
where $\Delta$ is the overdensity factor that defines the halos and $r_\Delta$ is its virial radius. Note that this mass does not represent the true subhalo mass since, as mentioned, subhalos suffer tidal forces. However, it is still a good proxy for their concentration, as tidal mass losses mainly affect the subhalo outskirts and, indeed, are not expected to change the inner structure significantly \citep{Kazantzidis:2003hb, Diemand:2008in}.

For an NFW DM density profile \citep{Navarro:1995iw, Navarro:1996gj}, 
\begin{equation}
\rho (r) = \frac{4 \, \rho_s}{(r/r_s) \, (1+r/r_s)^2} ~,
\end{equation}
where $r_s \equiv r_{-2}$ is the scale radius and $\rho_s$ is density at $r_s$. It can be shown that the relation between $c_{\rm V}$ and $c_\Delta$ is given by \citep{Diemand:2007qr}
\begin{eqnarray}
c_{\rm V}&=&\left( \frac{c_{\Delta}}{2.163}\right)^{3}\,\frac{f(R_{\rm{max}}/r_{s})}{f(c_\Delta)}\,\Delta ~,
\label{eq:cvcD}
\end{eqnarray}
where $f(x)= \mbox{ln}(1+x)-x/(1+x)$. Note that, since for an NFW profile $V_{\rm{max}}$ occurs at  $R_{\rm{max}}=2.163\, r_{s}$, the relation between both concentration definitions just depends on $\Delta$.

Furthermore, it is possible to rewrite the virial mass in terms of $R_{\rm{max}}$ and $V_{\rm{max}}$ in the following way:
\begin{equation}
m_{\Delta}= \frac{f(c_\Delta)}{f(2.163)} \, \frac{R_{\rm{max}} \,V_{\rm{max}}^{2}}{G} ~,
\label{eq:m200}
\end{equation}
with $G$ the gravitational constant. 

 {We recall that NFW profiles have been shown to represent a poor description of the actual DM distribution inside subhalos. Other DM density profiles have been proposed that more accurately describe the inner subhalo structure, e.g., \citep{Hayashi:2002qv, Kazantzidis:2003hb, Penarrubia:2010jk}. In general, these profiles are similar to NFW in the inner regions while they present much steeper slopes in the outskirts as a consequence of the removal of material due to tidal interactions. Yet, in this work we will assume NFW as a first good enough approximation for our purposes, and will leave the use of more accurate subhalo profiles for future work.\footnote{ {Adopting pure NFW instead of properly evolved, modified NFW for subhalos is expected to slightly overestimate the computation of the annihilation boost factor in section \ref{sec:boost}. However, as we will discuss later in that section, we introduce a truncation radius that effectively mimics, to a good extent, the mentioned more accurate profiles for subhalos.}}}

Below, we will investigate the dependence of the subhalo concentration on subhalo (would-be virial) mass and distance to the host halo center. We will do so for both definitions of the concentration, $c_{\rm V}$ and $c_{200}$, by making use of N-body simulation data. As for our notation, below we use capital (small) letters to refer to halos (subhalos) or the index $^h$ (no index) for halos (subhalos) otherwise.

\subsection{Description of the data sets}  
\label{sec:datasets}	

High-resolution N-body cosmological simulations are mandatory in order to study subhalo properties in great detail. Ideally, these simulations should resolve the subhalo internal structure accurately down to the innermost subhalo regions and should provide excellent subhalo statistics.  In our work, we have considered two N-body cosmological simulations of Milky-Way-size halos: VL-II \citep{Diemand:2008in} and ELVIS \citep{Garrison-Kimmel:2013eoa}. In both cases, present-day ($z = 0$) halo catalogs are available for public download\footnote{VL-II: \url{http://www.ics.uzh.ch/~diemand/vl/} \\ ELVIS: \url{http://localgroup.ps.uci.edu/elvis/}} and we use the results for $V_{\rm max}$ and $R_{\rm max}$. Note that one may also study halo substructure properties by making use of large-scale-structure simulations such as BolshoiP and MultiDark \citep{Prada:2011jf, Riebe:2011gp, Klypin:2014kpa, Rodriguez-Puebla:2016ofw}, which in turn would allow to learn about subhalo properties up to the largest (sub)halo masses. This is left for future work.

VL-II follows the growth of a Milky Way-size system in a $\Lambda$CDM universe from redshift 104.3 to the present time. The simulation employs just over one billion particles of mass $4100$ $M_{\odot}$ to model the formation of a $M$=$1.93$ x $10^{12}$ $M_{\odot}$ halo and its substructure, where the halo and subhalo masses are obtained assuming an overdensity of 200 relative to the \emph{mean} matter density of the Universe (or 47.6 with respect to the \emph{critical} density of the Universe at $z=0$).  {Both halos and subhalos in the high resolution region of the VL-II run were identified using the {\texttt 6DFOF} halo and subhalo finder \citep{Diemand:2006ey, Diemand:2006ik}}. More than $40000$ individual subhalos within the host halo are resolved within $R = 402$~kpc. Yet, the abundances and properties of many of these subhalos are affected by resolution effects and, as a result, the simulation team provides a reliable subsample of $\sim 9400$ subhalos with masses above $\sim 10^{6}~ M_{\odot}$.  VL-II adopted the cosmological parameters from the WMAP 3-year data release.

ELVIS contains 48 Milky-Way-size halos, of which half are in paired configurations, similar to the Milky Way and the Andromeda galaxy. The other half are isolated halos that are mass-matched to those in the pairs. In addition, high-resolution simulations of three isolated halos were performed. All simulations were initialized at redshift $z=125$. The mass resolution for the 48 galaxy-size halos is about $10^{5} \,M_{\odot}$, while the particle mass for the higher resolution set is $2.35\,\rm{x}\,10^{4} \,M_{\odot}$.   {Self-bound DM structures are identified with the {\texttt Rockstar} halo finder \citep{Behroozi:2011ju}, which was also used to compute the subhalo masses.} The virial mass of halos and subhalos is defined as the mass within the radius enclosing 97 times the \emph{critical} density of the Universe. The distribution of the virial masses of field halos covers the range $(1.0 - 2.85) \times 10^{12} \,M_{\odot}$. In addition, ELVIS resolves over $50000$ subhalos with masses above $\sim 10^6 M_{\odot}$. There is no statistical correlation among the field halos since they were extracted from independent collisionless simulations. Cosmological parameters were taken from WMAP 7-year results.

We provide a summary of the most relevant parameters of both simulations in Tab.~\ref{tab:param}. Let us note that the fact that $\Omega_{m}$ and $\sigma_8$ are lower for the WMAP 3-year than for the WMAP 7-year data set, implies that halos assemble later for WMAP 3-year cosmology (see, e.g., \citet{Maccio':2008xb}). However, the effect is expected to be small given the relatively close $\sigma_8$ values of both simulations and, indeed, as we show in the next section, we observe a very weak dependence of the concentration values on the cosmological parameters, both data sets being in good agreement with each other within their statistical dispersion. We also note that we present our results for $c_{\Delta}$ in the next section adopting $\Delta = 200$ as the value for the overdensity to define halos and subhalos. This is different from the $\Delta$ value used in each simulation, as described above and in Tab.~\ref{tab:param}, which implies that our $c_{\Delta}$ values are lower than those obtained if using the overdensities adopted in the simulations to define halos and subhalos. However, by doing so we are able to merge the results of both simulations and treat them on the same footing for our purposes.

\begin{table} 
	\begin{center}
		\begin{tabular}{| l c c c c c c c | }
			\hline
			& $\Omega_{\rm m,0}$ & $\Omega_\Lambda$ & $h$ & $n_s$ & $\sigma_{8}$ & $\Delta$ & $N_{\rm{sub}}$\\ \hline \hline
			VL-II   & 0.238 & 0.762 & 0.73 & 0.951 & 0.74  & 47.6 & 6914 \\ \hline 
			ELVIS & 0.266 & 0.734 & 0.71 & 0.963 & 0.801 & 97   &  35292 \\ \hline  
		\end{tabular}
	\end{center}
	\caption{VL-II and ELVIS most relevant parameters for this work. Columns 2--6 indicate the set of cosmological parameters used in each simulation; column 7 is the overdensity $\Delta$ over the critical density of the Universe; and column 8 denotes the number of subhalos, $N_{sub}$, that were finally used in our study (see Sec.~\ref{sec:fits} for further details). This number does not correspond to the actual number of subhalos present in the simulations, which is substantially larger.}
	\label{tab:param}
\end{table}

\subsection{Subhalo concentrations}
\label{sec:fits}
 
It is well known that subhalos exhibit concentrations that differ substantially from that of field halos of the same mass, the latter being found to be less concentrated \citep{Ghigna:1999sn, Bullock:1999he, Moore:1999nt, Ullio:2002pj, Diemand:2007qr, Diemand:2008in, Diemand:2009bm, Pieri:2009je, Bartels:2015uba}. Indeed, subhalos are subject to tidal forces that remove material from their outskirts, making them more compact. As a result, during this process $R_{\rm{max}}$ becomes smaller and the enclosed mean subhalo density, codified in $c_{\rm V}$ (Eq.~(\ref{eq:cvrho})), increases \citep{Diemand:2006ik, Kuhlen:2008aw, Springel:2008cc}.

In this section, we derive an accurate fit for the concentration-mass relation for subhalos using VL-II and ELVIS simulations data. Moreover, previous work has shown that the subhalo concentration depends not only on the mass of the subhalo but also on the distance to the center of its host halo \citep{Diemand:2008in, Pieri:2009je}. We quantify both dependences and propose a subhalo concentration parametrization that depends on these two quantities.\footnote{Other possible dependences, such as the one with the host halo mass \citep{Dooley:2014osa, Hellwing:2015upa}, is left for future work. It is not possible to investigate this dependence with our current data sets since all our host halos have comparable masses, of the order of $10^{12} \,M_{\odot}$.}

In our analysis of VL-II data, we adopted the restriction on the subhalo maximum circular velocity applied in \citet{Diemand:2008in}, i.e., only subhalos with $V_{\rm{max}} > 3$~km/s are included to avoid resolution issues in the determination of $c_{\rm V}$. A similar concern was addressed for ELVIS in the following way: for the three high-resolution halos, only subhalos with $V_{\rm{max}}>5$~km/s were considered, whereas all subhalos in the other 48 host halos were used ($V_{\rm{max}}>8$~km/s).

\begin{figure*}
	\centering	
	\includegraphics[width=0.49\textwidth]{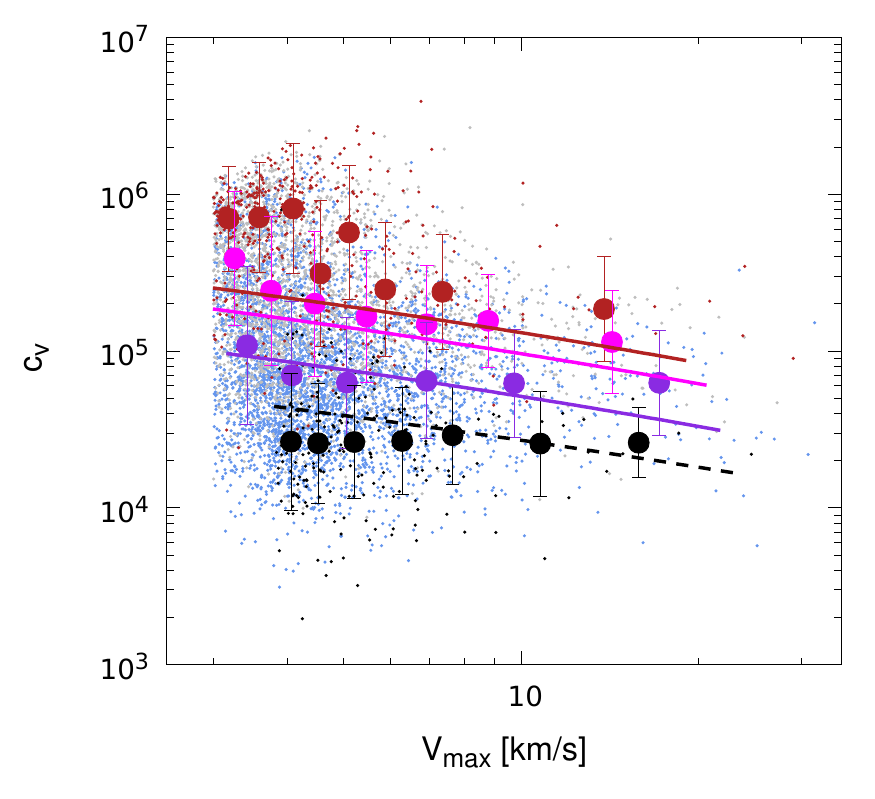}
	\includegraphics[width=0.49\textwidth]{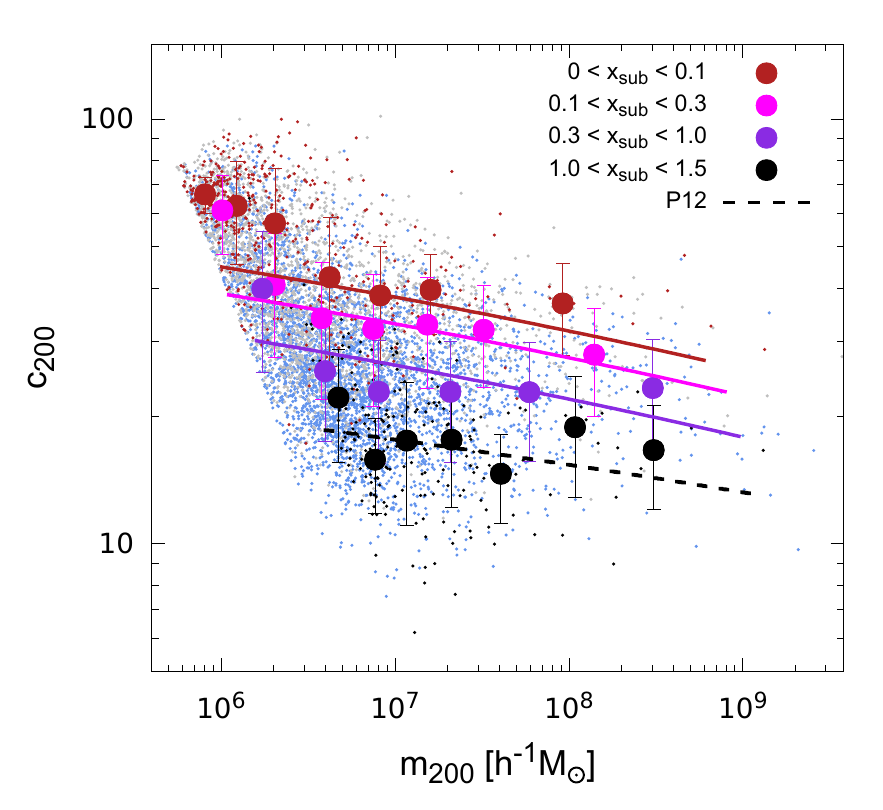} \\
	\includegraphics[width=0.49\textwidth]{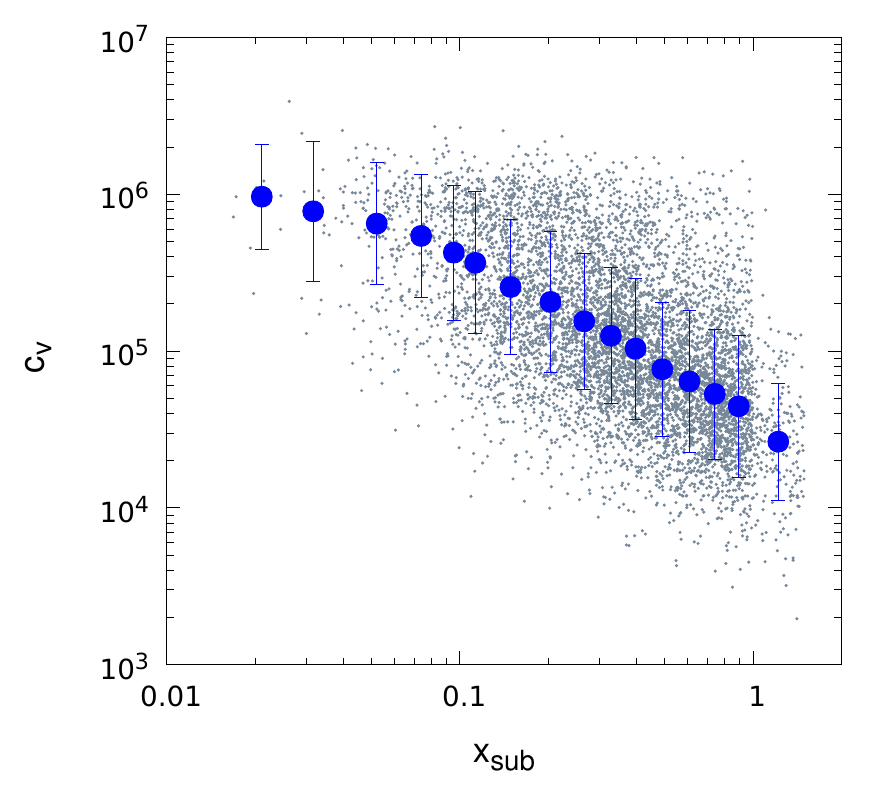}
	\includegraphics[width=0.49\textwidth]{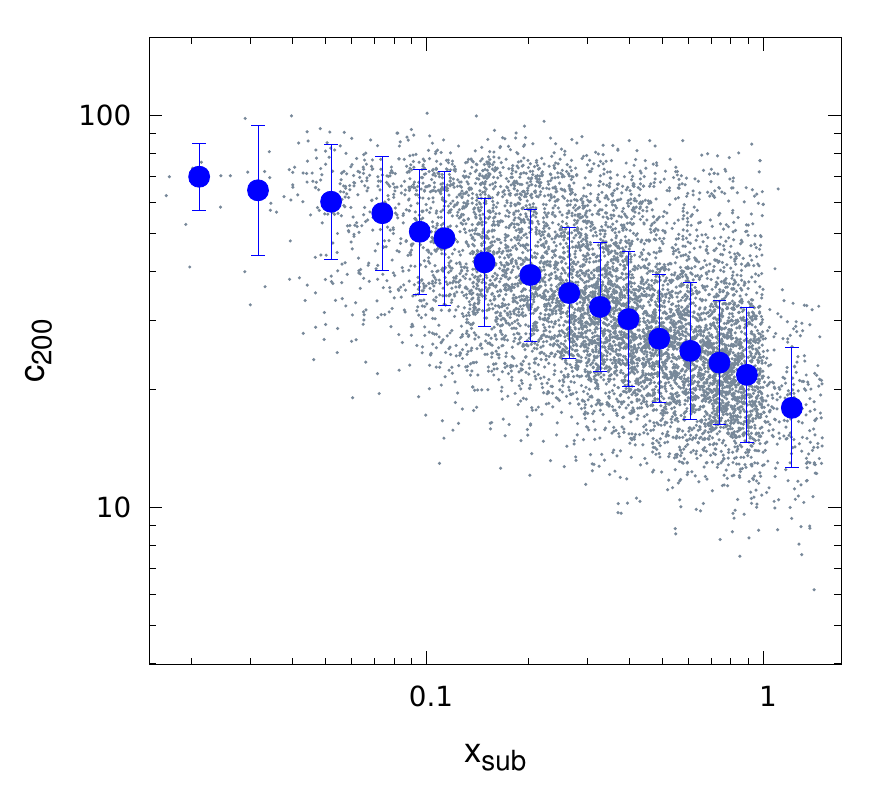}
	\caption{Median halo and subhalo concentrations and $1\sigma$ errors as found in the VL-II simulation \citep{Diemand:2008in}. The concentrations for all individual halos and subhalos are also shown (smaller dots in the background). {\it Top panels:} Results for subhalos depicted for three different bins of the distance to the center of the host halo. From top to bottom: bin I (red dots), II (magenta dots and gray background dots) and III (purple dots); see text for details. The black dots correspond to the halo median concentrations in the calibration bin beyond $R_{\Delta}$. The left panel shows the median $c_{\rm V}$ as a function of $V_{\rm{max}}$, while the right panel is for $c_{200}$ as a function of $m_{200}$. We also show the results of our fits (solid colored lines) and the P12 parametrization for the concentration of field halos (dashed black lines) \citep{Prada:2011jf} using the fit in \citet{Sanchez-Conde:2013yxa}. {\it Bottom panels:} Median $c_{\rm V}$ (left) and $c_{200}$ (right) as a function of the distance to the center of the host halo normalized to $R_{\Delta}$, $x_{\rm sub}$. All (sub)halo masses have been included in these two plots.}
	\label{fig:cv-c200-VLII} 
\end{figure*}

\begin{figure*}
	\centering 
	\includegraphics[width=0.49\textwidth]{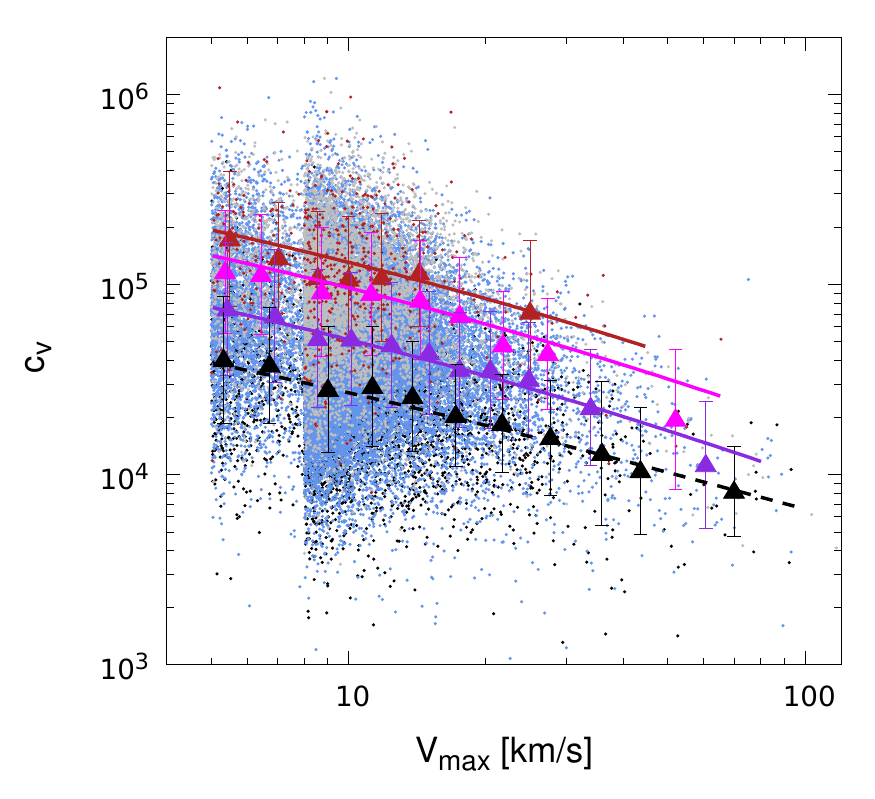}	
	\includegraphics[width=0.49\textwidth]{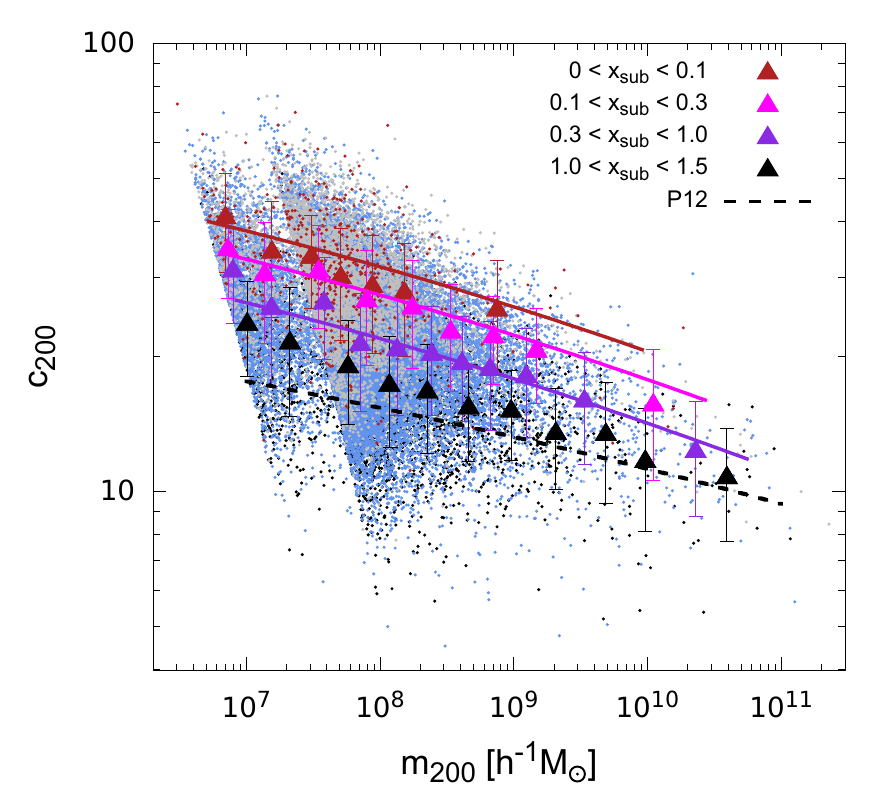} \\
	\includegraphics[width=0.49\textwidth]{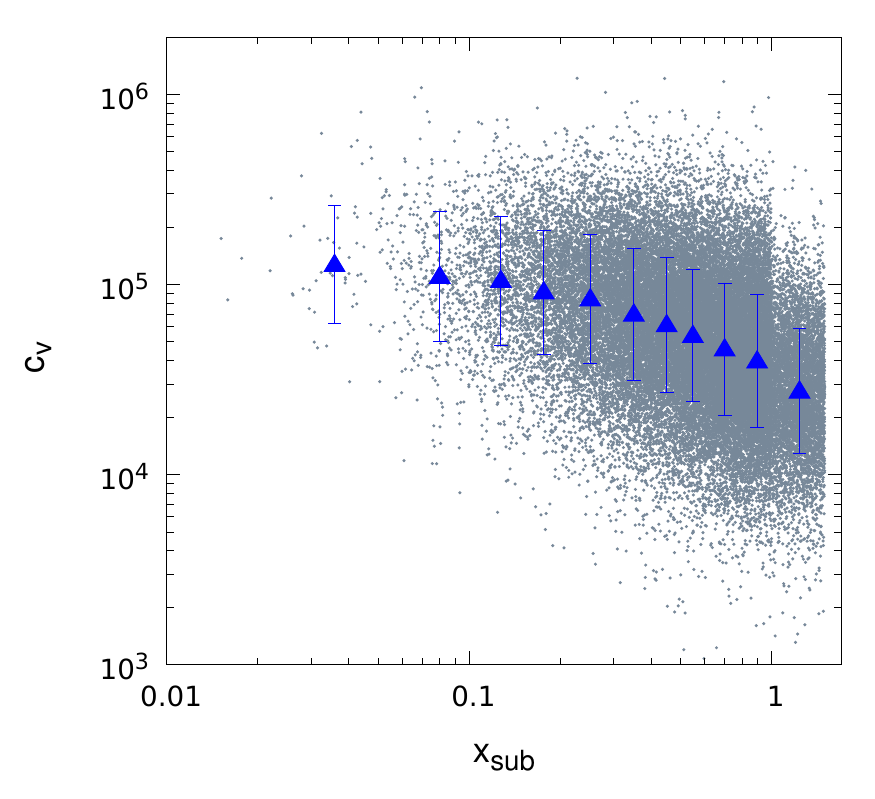}
	\includegraphics[width=0.49\textwidth]{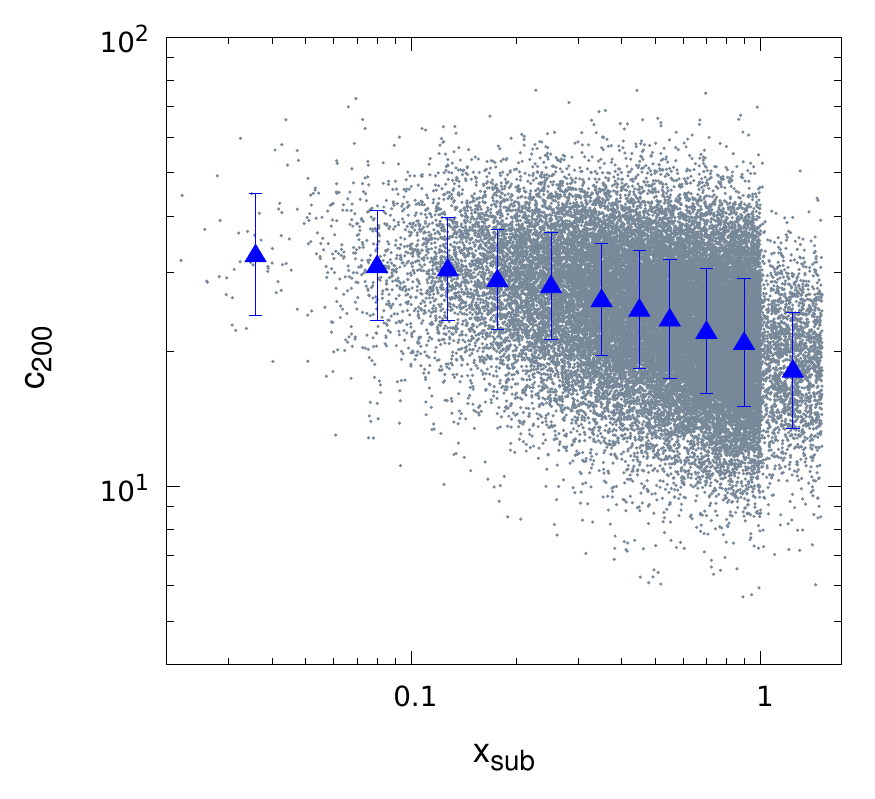}
	\caption{Same as Fig.~\ref{fig:cv-c200-VLII}, but for the ELVIS simulations \citep{Garrison-Kimmel:2013eoa}. The small dots forming the leftmost cloud correspond to the three halos with higher resolution in the simulation while the right cloud of points corresponds to the subhalos from the other 48 Milky-Way-size halos (see text).}
\label{fig:cv-c200-ELVIS}
\end{figure*}

For both simulations, we implemented three radial bins within the virial radius of the host halo, $R_{\Delta}$, as well as a calibration bin beyond the halo boundary. The innermost radial bin contains subhalos at a distance $R_{\rm sub}$ from the host halo center, $x_{\rm sub} \equiv R_{\rm sub}/R_{\Delta} < 0.1$ (bin I), while the second radial bin is defined as $0.1 < x_{\rm{sub}} < 0.3 $ (bin II), and the third bin contains subhalos within $0.3 < x_{\rm{sub}} < 1$ (bin III). In addition, a calibration bin has been included beyond $R_{\Delta}$ to estimate field halo concentrations and to compare those with subhalo concentrations. The calibration bin contains halos at distances to the host halo center up to $1.5\, R_{\Delta}$. In order to remove from our calibration bin any possible halo that was inside $R_{\Delta}$ in the past (and thus was a subhalo), we only consider structures such that the maximum circular velocity achieved over its entire existence falls within the $5\%$ of the velocity at $z=0$ \citep{Sanchez-Conde:2013yxa}. Then, for each radial bin, we have grouped subhalos in bins of $V_{\rm max}$ and have obtained the medians of $c_{\rm V}$. Different bin sizes were chosen to cover the entire  $V_{\rm max}$ range with a similar number of subhalos per bin. In the top-left panels of Figs.~\ref{fig:cv-c200-VLII} and~\ref{fig:cv-c200-ELVIS}, we show the median $c_{\rm V}(V_{\rm{max}})$ values and $1\sigma$ errors found for VL-II and ELVIS, respectively.

The same analysis in radial bins was performed for $c_{200}$, which was found by applying the $c_{\rm V} - c_{200}$ relation of Eq.(\ref{eq:cvcD}) to the $c_{\rm V}(V_{\rm{max}})$ values found for every subhalo in the simulations. In this case, the medians were obtained considering several mass bins. 
 {We recall that Eq.(\ref{eq:cvcD}) implicitly assumes NFW profiles for subhalos which, as already said, do not represent an accurate description of the actual subhalo inner structure. Thus, the $c_{200}$ values so derived must be taken carefully and mainly for comparison purposes: strictly speaking, they correspond to the $c_{200}$ values that non-truncated halos would exhibit for the corresponding value of $c_{\rm V}$ considered (i.e., $V_{\rm max}$ and $R_{\rm{max}}$ pair of values).}
The top-right panels of Figs.~\ref{fig:cv-c200-VLII} and~\ref{fig:cv-c200-ELVIS} show the median $c_{200}(m_{200})$ values and $1\sigma$ errors for VL-II and ELVIS, respectively, as well as the $c_{200}(m_{200})$ relation for field halos derived by \citet{Prada:2011jf} from the analysis of the Bolshoi and MultiDark simulations. In the top panels, we also depict the results of our fits (cf. Eqs.~(\ref{eq:c200-fit})-(\ref{eq:cv-cal})).

It is also interesting to explicitly show the radial dependence of the subhalo concentration, which is very similar for all subhalo masses. The bottom panels of Figs.~\ref{fig:cv-c200-VLII} and~\ref{fig:cv-c200-ELVIS} depict, for VL-II and ELVIS, respectively, the medians and $1\sigma$ errors for $c_{\rm V}$ and $c_{200}$ as a function of the distance from the center of the host halo in units of $R_{\Delta}$, both for subhalos within $R_{\Delta}$ and for the halos in the calibration bin, ($x_{\rm sub} > 1$).\footnote{Note that the bottom-left panel in Fig.~\ref{fig:cv-c200-VLII} is very similar to the bottom panel of the supplementary Fig.~4 in the VL-II paper \citep{Diemand:2008in}.} These bottom panels cover the entire range of each simulation, $m_{200} \sim (10^6 - 10^{9})~h^{-1} M_{\odot}$ in VL-II and $m_{200} \sim (10^7 - 10^{11})~h^{-1} M_{\odot}$ in ELVIS.

As can be seen from Figs.~\ref{fig:cv-c200-VLII} and~\ref{fig:cv-c200-ELVIS}, the median subhalo concentration increases towards the center of the host halo and is significantly larger than that of field halos, in good agreement with \citet{Diemand:2007qr, Diemand:2008in}. More precisely, we find that $c_{200}$ subhalo values can be almost a factor 3 larger than those of field halos of the same mass (for the innermost radial bin and the less massive subhalos in VL-II), and it is typically between a factor $\sim1.5-2$ (its exact number depending on the subhalo mass and distance to the host halo center). For ELVIS, the ratio between subhalo and halo $c_{200}$ values is typically lower compared to VL-II for the same subhalo mass and distance to the host halo center and, indeed, never reaches a factor 2. In terms of $c_{\rm V}$, these subhalo-to-halo concentration ratios are between a factor $\sim2-14$ and $\sim2-5$ for VL-II and ELVIS, respectively, in the $V_{\rm{max}}$ range where they overlap, i.e., $V_{\rm{max}} \sim (6 - 20)$~km/s. 

\begin{figure*}
	\begin{tabular}{c}
		\includegraphics[width=0.98\textwidth]{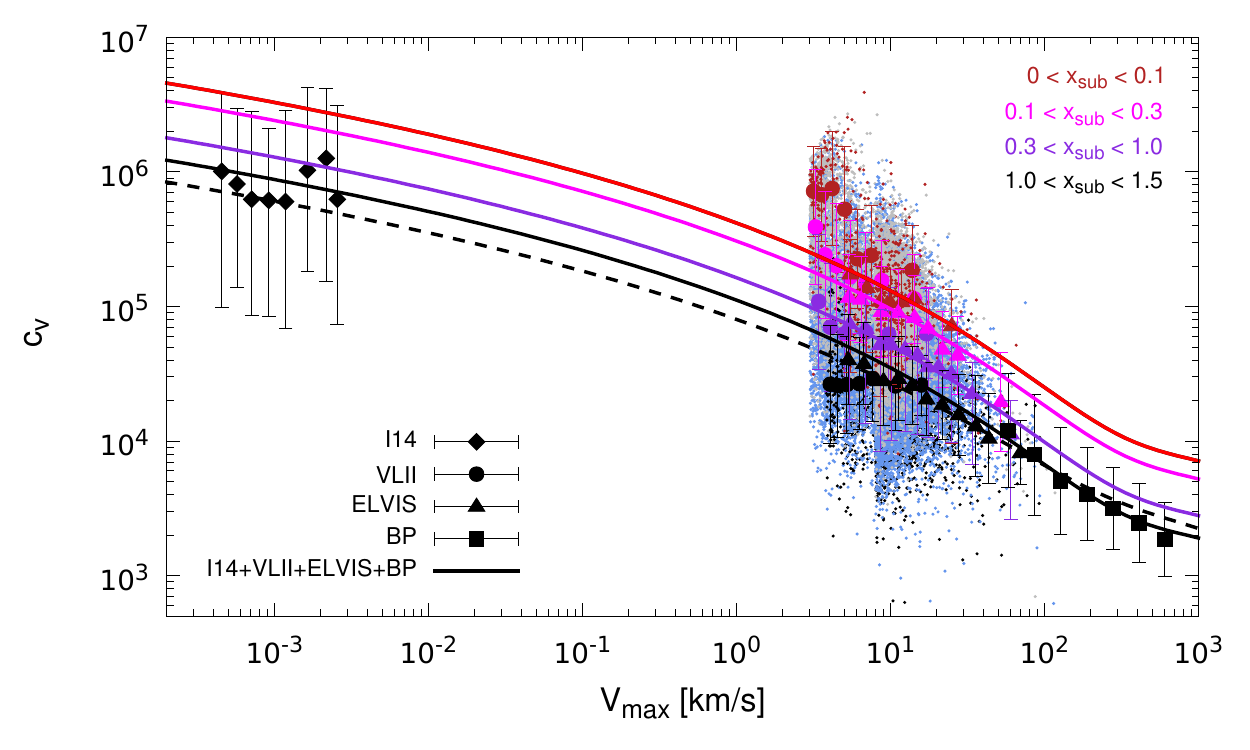} \\
		\includegraphics[width=0.98\textwidth]{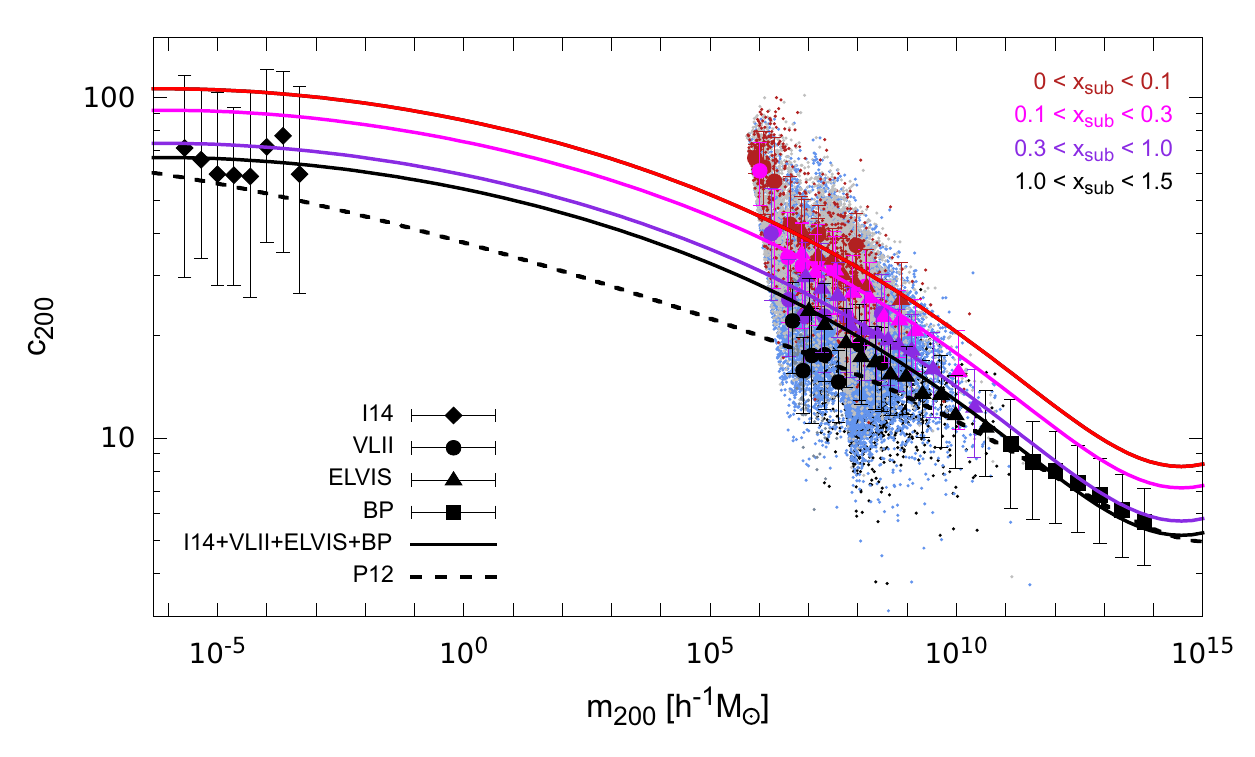} 
	\end{tabular}
	\caption{{\it Top panel}: Median halo (black symbols) and subhalo (colored symbols, for each radial bin) concentration parameter $c_{\rm V}$, and $1\sigma$ errors, as a function of $V_{\rm max}$ as found in the VL-II (circles) and ELVIS simulations (triangles). The concentrations for all individual subhalos are also shown (smaller dots in the background). The results for  {field} microhalos from I14 \citep{Ishiyama:2014uoa} and for more massive  {main} halos from BolshoiP (BP) \citep{Klypin:2014kpa, Rodriguez-Puebla:2016ofw} are shown by black diamonds and squares, respectively. We also show our fits for halos given by Eq.~(\ref{eq:cv-cal}) (dashed black line) and subhalos in Eq.~(\ref{eq:cv-fit}) (solid colored lines) for each of the three radial bins considered. {\it Bottom panel}: Same as top panel, but for $c_{200}$ as a function of $m_{200}$. Our proposed fit for each of the radial bins, Eq.~(\ref{eq:c200-fit}), and the P12 parametrization for the concentration of halos \citep{Prada:2011jf} using the fit obtained in \citet{Sanchez-Conde:2013yxa}, are also shown.} 	
	\label{fig:ELVIS-VLII}
\end{figure*}

For the sake of comparison, in the top panel of Fig.~\ref{fig:ELVIS-VLII}, we show the median $c_{\rm V}(V_{\rm{max}})$ relation and the $1\sigma$ errors found from the results of the VL-II and ELVIS simulations together. They cover the range $V_{\rm{max}} \sim (3 - 100)$~km/s. Likewise, the median $c_{200}(m_{200})$ relation and the $1\sigma$ errors for both simulations are also shown in the bottom panel of Fig.~\ref{fig:ELVIS-VLII}. Altogether, they cover the subhalo mass range $m_{200} \sim (10^6 - 10^{11})~h^{-1} M_{\odot}$. It can be seen that there is an excellent agreement between the results from the VL-II and ELVIS data within errors.

We now provide parametrizations for the median subhalo $c_{\rm V} (V_{\rm max}, x_{\rm sub})$ and $c_{200}(m_{200}, x_{\rm sub})$ relations, based on the results obtained from both VL-II and ELVIS. In order to reduce the uncertainties when extrapolating outside the range probed by these simulations, we also use the concentration of more massive  {main} halos obtained from the BolshoiP simulation \citep{Klypin:2014kpa, Rodriguez-Puebla:2016ofw} and  {of main halos with masses in the range $(10^{-6} - 10^{-3}) \, h^{-1} \, M_\odot$} obtained by \citet{Ishiyama:2014uoa}. Formally, the way we add these halos in our subhalo fit is by assuming the subhalo concentration at $x_{\rm sub} = 1$ to be equal to the concentration of halos of the same mass. All these points from the simulations, as well as our fits are shown in both panels of Fig.~\ref{fig:ELVIS-VLII}.

For the case of the median subhalo concentration-$V_{\rm max}$ relation, $c_{\rm V}(V_{\rm max}, x_{\rm sub})$, we get
\begin{eqnarray}
c_{\rm V}(V_{\rm max},x_{\rm sub}) & = & c_{0} \, \left[1+\sum_{i=1}^{3} \,  \left[a_{i} \, \log\left(\frac{V_{\rm max}}{10 \, {\rm km/s}}\right)\right]^{i} \right] \times \nonumber \\ 
& & \left[1 +  b \, \log\left(x_{\rm sub}\right) \right] ~,
\label{eq:cv-fit}
\end{eqnarray}
where $c_{0} = 3.5 \times 10^4$, $a_{i}=\left\{-1.38, \, 0.83, \, -0.49 \right\}$ and $b = -2.5$.  {In the top panel of Fig.~\ref{fig:ELVIS-VLII}, we show the results of our fit together with the median concentration values from both VL-II and ELVIS simulations, for all the radial bins considered in our work.} This fit works well for $10^{-4} \, {\rm km/s} \, \lesssim V_{\rm max} \lesssim 10^3$~km/s.

Likewise, we obtain a parametrization for $c_{200}$ as a function of $m_{200}$ and $x_{\rm sub}$ for subhalos:\footnote{Note that this parametrization diverges logarithmically for $x_{\rm sub} \rightarrow 0$; yet, this does not have any impact on the calculation of the boost factor in the next section.}
\begin{eqnarray}
c_{200}(m_{200},x_{\rm sub}) & = & c_0 \, \left[1+\sum_{i=1}^{3} \,  \left[a_{i} \, \log\left(\frac{m_{200}}{10^8 \, h^{-1} \, M_{\odot}}\right)\right]^{i} \right]
\times \nonumber \\ 
& & \left[1 +  b \, \log\left(x_{\rm sub}\right) \right] ~,
\label{eq:c200-fit}
\end{eqnarray}
with $c_0 = 19.9$, $a_{i}=\left\{-0.195, \, 0.089, \, 0.089 \right\}$ and $b = -0.54$. This parametrization agrees well with expectations for field halos beyond $R_{\Delta}$ and, in particular, with P12. We note again that it implicitly assumes an NFW density profile for subhalos, following the discussion in Sec.~\ref{sec:definitions}. 

Like for the case of $c_{\rm V}$, in the bottom panel of Fig.~\ref{fig:ELVIS-VLII}, we show the results of our fit together with the median concentration values from both VL-II and ELVIS simulations, for all the radial bins considered in our work. It works well in the subhalo mass range $10^{-6} \, h^{-1} \, M_\odot \lesssim m_{200} \lesssim 10^{15} \, h^{-1} \, M_\odot$.

In order to compute the boost factor in Sec.~\ref{sec:boost} we also need to have the concentration for the field halos. In the case of $c^h_{200}$ we will use the P12 parametrization. When using $c^h_{\rm V}$ we have no parametrization for field halos and only have information for subhalos. Nevertheless, as we discussed above, the concentration in the calibration bin agrees very well with the concentration of field halos, so we use these results along with the concentration of the more massive  {main} halos from the BolshoiP simulation \citep{Klypin:2014kpa, Rodriguez-Puebla:2016ofw} and that of  {field} microhalos from \citet{Ishiyama:2014uoa}. In order to compute $c^h_{\rm V}$ from $c^h_{200}$ in P12, we assume an NFW profile and use Eq.~(\ref{eq:cvcD}). Therefore, and analogously to subhalos, we obtain a fit for $c^h_{\rm V}$ for field halos. It is given by
\begin{equation}
c_{\rm V}^h(V^h_{\rm max}) = c_{0} \, \left[1+\sum_{i=1}^{3} \,  \left[a_{i} \, \log\left(\frac{V^h_{\rm max}}{10 \, {\rm km/s}}\right)\right]^{i} \right] ~,
\label{eq:cv-cal}
\end{equation}
where $c_{0} = 2.7 \times 10^4$ and $a_{i}=\left\{-1.26, \, 0.78, \, -0.47 \right\}$. This fit works well for $10^{-4} \, {\rm km/s} \, \lesssim V^h_{\rm max} \lesssim 10^3$~km/s.
 
The best-fit values for the three parametrizations of the concentration described above are indicated in Tab.~\ref{tab:fits}.

\begin{table}
	\begin{center}
		\begin{tabular}{| l  c  c  c  c  c | }
			\hline
			& $c_0$ & $a_1$ & $a_2$ & $a_3$ & $b$ \\ \hline \hline 
			$c_{\rm V} \,$  [Eq.~(\ref{eq:cv-fit})]       & $3.5 \times 10^4$ & -1.38 & 0.83 & -0.49  & -2.5 \\ \hline 
			$c_{200} \,$ [Eq.~(\ref{eq:c200-fit})]     & 19.9                & -0.195 & 0.089 & 0.089  & -0.54 \\ \hline 			
			$c^h_{\rm V} \,$  [Eq.~(\ref{eq:cv-cal})] & $2.7 \times 10^4$ & -1.26 & 0.78 & -0.47 &  $-$ \\ \hline 
		\end{tabular}
	\end{center}
	\caption{Best-fit values of the parametrizations for the concentration parameter for subhalos as a function of the radial distance $x_{\rm sub}$ and as a function of $V_{\rm max}$ ($c_{\rm V}$) and $m_{200}$ ($c_{200}$), and in the calibration bin as a function of $V_{\rm max}$, i.e., for halos ($c^h_{\rm V}$).}
	\label{tab:fits}
\end{table}

\subsection{Scatter of the subhalo concentration}
\label{sec:scatter}

\begin{figure*}
	\centering
	\includegraphics[width=1.1\textwidth]{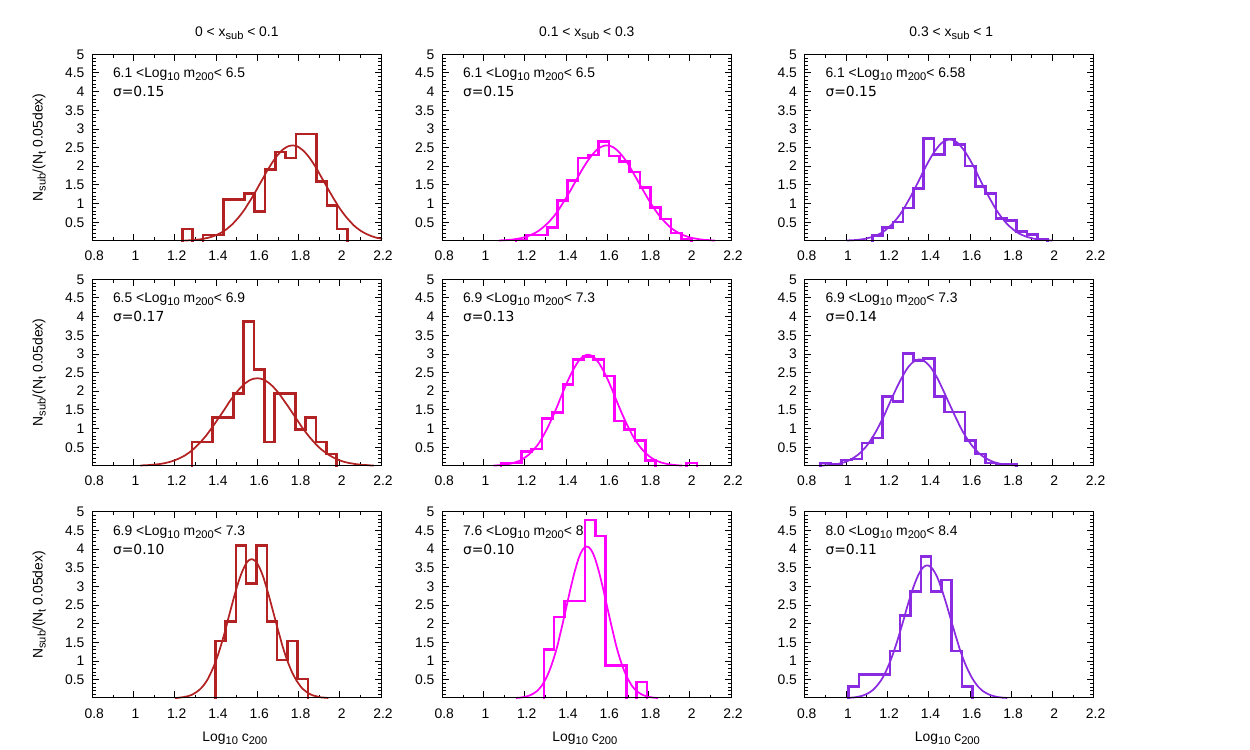}
	\caption{Scatter of the $c_{200}$ concentration parameter as measured in the VL-II simulation \citep{Diemand:2008in}, for different mass bins (rows) and the three radial bins considered in this work (columns). The smooth thin lines overimposed to the histograms correspond to the fits to the data using the log-normal distribution function given by Eq.~(\ref{eq:scatter}). We also indicate the value of the scatter, $\sigma_{\log_{10} c_{200}} \equiv \sigma$, in each bin.}
	\label{fig:sa-VLII}
\end{figure*}

\begin{figure*}
	\centering
	\includegraphics[width=1.1\textwidth]{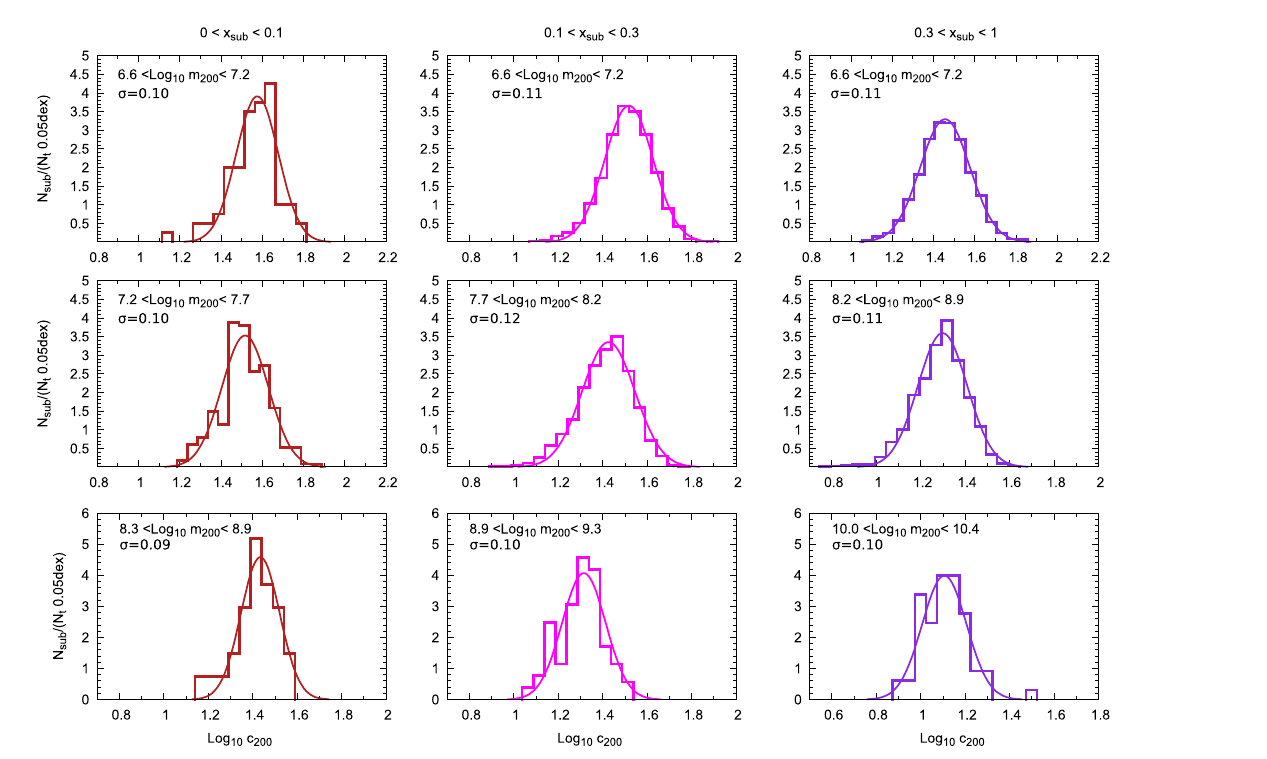}
	\caption{Same as Fig.~\ref{fig:sa-VLII} but for the ELVIS simulation \citep{Garrison-Kimmel:2013eoa}.}
	\label{fig:sa-ELVIS}
\end{figure*}

Due to the  {randomness of the initial fluctuations, which give rise to large variance in the assembly histories of DM halos}, an intrinsic, irreducible scatter is expected on the concentration of DM halos of the same mass \citep{Jing:1998xj, Bullock:1999he, Maccio':2006nu, Neto:2007vq, Maccio':2008xb, Dutton:2014xda}, which mainly reflects their slightly different formation epochs \citep{Wechsler:2001cs}. The size of this \emph{halo-to-halo} scatter is an important property of the underlying cosmological model that, ideally, could be used as an additional test for the model itself. In simulations, the measurement of the scatter of the halo concentration is also subject to other non-intrinsic effects that are expected to enlarge it, such as Poisson noise due to the limited number of halos available (especially close to the high-mass end of the halo mass range proved by the simulation, where halos are more scarce) and poor resolution to properly resolve the internal structure of the smallest halos and thus to derive their concentrations, e.g., \citep{Bullock:1999he}. 

For the case of subhalos, the intrinsic \emph{subhalo-to-subhalo} scatter has been claimed to be larger than the halo-to-halo scatter in field halos. A more complex formation history and other physical phenomena not present or negligible in the case of distinct halos, such as interactions or stripping, has been proposed to be the source of this larger value of the scatter for subhalos \citep{Bullock:1999he}.

Here, we investigate the scatter of the subhalo concentrations in both VL-II and ELVIS, for the three different radial bins we have considered. In order to do so, we also group the data in several mass bins. As for the case of field halos, we consider a log-normal distribution \citep{Jing:1998xj}:\footnote{See, however, \citet{Reed:2010gh, Bhattacharya:2011vr, Diemer:2014gba}.}
\begin{eqnarray}
P(c_{200}) = \frac{1}{c_{200} \ln 10 \sqrt{2 \pi} \, \sigma_{\log_{10} c_{200}}} \,  
e^{- \frac{1}{2} \, \left(\frac{\log_{10} c_{200}-\log_{10} c_{200,0}}{\sigma_{\log_{10} c_{200}}}\right)^2} ~,
\label{eq:scatter}
\end{eqnarray}
where $\sigma_{\log_{10} c_{200}}$ is the scatter and $\log_{10} c_{200,0}$ is the median, which is what we have obtained and discussed in the previous subsection.

The results are shown in Figs.~\ref{fig:sa-VLII} and~\ref{fig:sa-ELVIS}, where we depict the probability distributions of $\log_{10} c_{200}$ for different mass intervals in each of the radial bins considered, for VL-II and ELVIS, respectively. In each of the panels we indicate the mass and radial bin we consider, and indicate the scatter of the subhalo concentration parameter found in that bin. As we can see, the scatter does not depend on the subhalo position within the host halo, and only a very weak dependence with the mass, not statistically significant, is observed in VL-II (Fig.~\ref{fig:sa-VLII}), which is not observed in ELVIS (Fig.~\ref{fig:sa-ELVIS}).  {All in all, we stress again that the results from both simulations are in excellent agreement within errors}. We obtain a typical subhalo-to-subhalo scatter of $\sigma_{\log_{10} c_{200}} = 0.13$ for VL-II and $\sigma_{\log_{10} c_{200}} = 0.11$ for ELVIS. 
 {Interestingly, this is similar to what we find for field halos beyond the virial radii of both VL-II and ELVIS, and comparable to what it has been found for field halos in previous studies as well \citep{Maccio':2006nu, Neto:2007vq, Maccio':2008xb, Dutton:2014xda}. }

 {This result seems to contradict the idea of subhalos exhibiting a significantly larger scatter than field halos. The apparent discrepancy though could mainly be due to the fact that our scatter calculation was performed in radial bins, which intentionally group subhalos according to more similar formation histories among them, thus yielding a lower concentration scatter in this case. Indeed, when deriving the ``global'' scatter of the entire subhalo population for a given mass bin independently of the subhalo location within the host, we found a slight yet clear increase of the scatter up to $\sigma_{\log_{10} c_{200}} \sim 0.15$ for VL-II and $\sigma_{\log_{10} c_{200}} \sim 0.13$ for ELVIS. These values are slightly larger than those for field halos in both simulations and thus, in line with the general idea of subhalo concentrations exhibiting larger scatter. Yet, we note that this increase seems significantly more moderate with respect to what it has been traditionally assumed. The concise reasons for such a result and its implications deserve further study, which will be done elsewhere.} 
Note as well, that the found values probably represent a conservative, upper limit estimate of the intrinsic scatter for subhalos in these simulations, since we did not model and subtract any of the other possible sources of (spurious, non physical) scatter in the data that were mentioned above.


\section{Halo substructure boost factor for dark matter annihilation}  
\label{sec:boost}

\subsection{Context}  
\label{sec:boostcontext}

One of the most active lines of research in DM physics is that of indirect searches of the DM annihilation products (gamma-rays, neutrinos and antimatter) in different astrophysical regions \citep{Conrad:2015bsa}. The flux of this potential DM signal is proportional to the square of the DM density and thus, regions where the DM density is higher are, in principle, the most promising targets. 

Halo substructure is expected to play a relevant role on this search. On one hand, some of the closest subhalos may represent good targets by themselves \citep{Baltz:1999ra, Tasitsiomi:2002vh, Koushiappas:2003bn, Diemand:2005vz, Baltz:2006sv, Pieri:2007ir, Kuhlen:2008aw, Buckley:2010vg, Belikov:2011pu, Ackermann:2012nb, Mirabal:2012em, Zechlin:2012by, Berlin:2013dva, Bertoni:2015mla, Schoonenberg:2016aml}. On the other hand, hierarchical structure formation in the $\Lambda$CDM paradigm implies that larger halos are teeming with smaller subhalos and this clumpy distribution is expected to significantly boost the DM annihilation signal \citep{Bergstrom:1998jj, CalcaneoRoldan:2000yt, Aloisio:2002yq, Stoehr:2003hf, Koushiappas:2003bn, Lavalle:1900wn, Kuhlen:2008aw, Charbonnier:2011ft, SanchezConde:2011ap, Pinzke:2011ek, Gao:2011rf, Nezri:2012tu, Anderhalden:2013wd, Zavala:2013lia, Sanchez-Conde:2013yxa, Ishiyama:2014uoa, Zavala:2015ura}. In this section, we quantify the latter and discuss about the implications a precise determination of the internal properties of subhalos has for indirect DM searches. 

During the last decade, numerous efforts have being devoted to estimate the extent of this DM annihilation flux enhancement due to the presence of halo substructure. Several analytical approaches based on the extended Press-Schechter theory have addressed the properties of the subhalo population in larger halos (see, e.g., \citet{Pieri:2007ir, Giocoli:2007gf, Giocoli:2008qg, Pieri:2008nb}) and used results from early N-body cosmological simulations about the distribution of subhalos in host halos \citep{Diemand:2005vz, Diemand:2005rd} to estimate the impact on gamma-ray searches. Also, more refined simulations in the past years have allowed us to better characterize the population and the structure of DM subhalos \citep{Kuhlen:2008aw, Springel:2008cc, Diemand:2008in, Hellwing:2015upa, Rodriguez-Puebla:2016ofw} and different algorithms have been proposed in order to identify small structures within simulated host halos \citep{Ghigna:1999sn, Weller:2004sc, Shaw:2005dy, Springel:2000qu, Gill:2004ki, Gill:2004km, Tormen:2003kb, Giocoli:2007uv, Giocoli:2009ie, Han:2011ad, Behroozi:2011ju}. However, the vast majority of this simulation work has been done for galaxy- and cluster-sized DM halos and, indeed, although a few simulations of  {(field)} microhalos are available \citep{Diemand:2005vz, Anderhalden:2013wd, Ishiyama:2011af, Ishiyama:2014uoa}, simulating substructure down to the smallest predicted halo masses and up to the present time remains very challenging. Yet, in order to estimate the boost factor due to DM annihilations, the contribution from the smallest halos needs to be included. The minimum halo mass depends on the free-streaming of DM particles from high to low density regions \citep{Zybin:1999ic, Hofmann:2001bi, Berezinsky:2003vn, Green:2005fa} and on the effect of acoustic oscillations \citep{Loeb:2005pm, Bertschinger:2006nq}. These processes depend on the particle physics and cosmological models \citep{Profumo:2006bv, Bringmann:2006mu, Bringmann:2009vf, vandenAarssen:2012ag, Cornell:2012tb, Gondolo:2012vh, Cornell:2013rza, Shoemaker:2013tda, Diamanti:2015kma} and hence, the minimum mass is very uncertain,\footnote{In the case of subhalos, additional violent processes might be also at work during their accretion and merging into larger halos that could significantly alter the subhalo survival probability and set, in practice, a minimum subhalo mass at present time different to that of host halos (see, e.g., \citet{Zhao:2005mb, Berezinsky:2005py, Goerdt:2006hp}).} with possible values within $M_{\rm min}=10^{-12}-10^{-4}M_\odot$.  In what follows, we set it to $M_{\rm min} = 10^{-6} \, M_\odot$.
 
The luminosity from DM annihilations in a halo or subhalo scales as the third power of the concentration. Thus, the substructure boost is very sensitive to the way the subhalos' internal structure is modeled. Moreover, since smaller (sub)halos possess larger concentrations, and are much more numerous, they are expected to dominate in the computation of the boost. However, as mentioned above, there is a lack of simulations at small halo scales (see, e.g., the right panel of Fig.~1 in \citet{Sanchez-Conde:2013yxa}) so only extrapolations over many orders of magnitude in halo mass are possible.\footnote{We note that the  {field} microhalo concentration results by \citet{Ishiyama:2014uoa}, although outstanding and extremely important in this context, were obtained at very high redshifts, so extrapolations of such concentrations down to the present time were performed in Fig.~1 of \citet{Sanchez-Conde:2013yxa}. Thus, strictly speaking, we still lack simulations that track the formation and evolution of the smallest halos all the way down to $z=0$.} Some works have simply extrapolated the power-law behavior of the concentration observed above the simulation resolution limit all the way down to the minimum halo mass. However, these power-law extrapolations to low masses, which tend to predict very large boost factors \citep{Pinzke:2011ek, Gao:2011rf}, are at odds with recent results on microhalo simulations \citep{Ishiyama:2014uoa} and are not expected either in the $\Lambda$CDM cosmological model \citep{Prada:2011jf, Sanchez-Conde:2013yxa, Ng:2013xha}. Indeed, a flattening of the concentration towards low halo masses is naturally expected in $\Lambda$CDM: halo concentration is set by the halo formation time, which is nearly the same for a broad range of halo masses in the low-mass regime as a consequence of the power spectrum of matter fluctuations. Therefore, the natal concentrations of these small halos are also expected to be nearly the same, and so they will be at the current epoch. When taking into account the correct behavior of the concentration at small halo masses, moderate values of the substructure boost factor are found \citep{Ishiyama:2014uoa, Sanchez-Conde:2013yxa}.

 \subsection{Computation of the boost}  
\label{sec:boostcomput}

Previous work has traditionally computed the substructure boost to the DM annihilation signal by assuming the concentration of subhalos to be the same as the one of field halos of the same mass. This represents a fair first order approximation but, as we showed in the previous section, the subhalo concentrations can differ substantially from that of field halos. Here, we recompute the boost factor taking advantage of that gained from our studies of subhalo internal properties. We note that, in principle, the self-similarity of halo hierarchy implies that one should consider several levels of substructure in the calculation of the boost. Nevertheless, as has been shown \citep{Strigari:2006rd, Martinez:2009jh, Sanchez-Conde:2013yxa}, only counting down to the second level (i.e., sub-substructure) is necessary in practice.

The substructure boost factor, $B(M)$, is given by \citep{Strigari:2006rd, Kuhlen:2008aw}:\footnote{Note that this is not the boost factor for a solid angle along a given line of sight in our Milky Way, but it is the boost factor for halos which are \emph{fully contained} within the solid angle of observation and thus, are relatively distant from us.  {Moreover, it is well known that subhalo mass functions have an exponential cut-off at high masses. However, due to the steep fall of the subhalo mass function with mass, the precise value of the cut-off at the high-mass end has very little impact on our results. The correction due to cutting the integral at 10\% of the host halo mass is at the percent level or below (being larger for smaller halos). The same arguments apply for the subhalo $V_{\rm max}$ function described below.}}
\begin{eqnarray}
B(M) & = &  \frac{4 \, \pi \, R_{200}^3}{{\cal L}_{\rm smooth}(M)} \, \int_{M_{\rm min}}^{M} dm \, \int_{0}^{1} dx_{\rm sub} \nonumber \\
& & \frac{dn(m,x_{\rm sub})}{dm} \,{\cal L}(m,x_{\rm sub}) \, x_{\rm sub}^2   ~,
\label{eq:BMr}
\end{eqnarray}
where ${\cal L}_{\rm smooth}(M)$ is the luminosity from the smooth DM distribution (no substructures) of a halo of mass $M$, ${\cal L}(m,x_{\rm sub})$ is the luminosity of a subhalo of mass $m$ at a distance $R_{\rm sub}$ ($x_{\rm sub} = R_{\rm sub}/R_{200})$ from the center of the host halo and $dn(m,x_{\rm sub})/dm$ is the subhalo mass function per unit of volume. Defined in this way, a boost factor $B=0$ represents the case of no substructure.

The luminosity of a field halo of mass $M$ from the smooth distribution, ${\cal L}_{\rm smooth}(M)$, is defined as
\begin{eqnarray}
{\cal L}_{\rm smooth}(M) & \equiv & \int_0^{R_{200}} \rho_{\rm host}^2(r) \, 4 \, \pi \, r^2 \, dr = \\
& & \frac{M \,c_{200}^{h}(M)^3}{\left[f(c^h_{200}(M))\right]^{2}} \frac{200 \, \rho_c}{9} \, \left(1-\frac{1}{(1+c^h_{200}(M))^3} \right) ~, \nonumber
\label{eq:LM}
\end{eqnarray}
where in the last step we have assumed an NFW profile and for halos, we use the parametrization for the concentration parameter from \citet{Prada:2011jf} using the fit obtained in \citet{Sanchez-Conde:2013yxa}.

With this at hand, the luminosity of a subhalo of mass $m$ at a distance $R_{\rm sub}$ from the center of the host halo, ${\cal L}(m,x_{\rm sub})$, is defined as
\begin{equation}
{\cal L}(m,x_{\rm sub}) = \left[1+B(m,x_{\rm sub})\right] {\cal L}_{\rm smooth}(m,x_{\rm sub}) ~.
\end{equation}
where now ${\cal L}_{\rm smooth}(m,x_{\rm sub})$ is the luminosity for the smooth distribution of the given subhalo and $B(m,x_{\rm sub})$ is the boost factor due to the next level of substructure. The luminosity of a subhalo (sub-subhalo) is given by the same functional form as that of a field halo, but including the dependence of the concentration parameter on the position of the subhalo (sub-subhalo) inside the host halo (subhalo). 

In addition to the mentioned dependences, we note that subhalos are not homogeneously distributed within the host halo \citep{Springel:2008cc, Hellwing:2015upa, Rodriguez-Puebla:2016ofw}. However,  {by using a similar radial distribution as that in \citet{Springel:2008cc, Hellwing:2015upa},} we have checked that the precise spatial distribution of subhalos inside halos has only a relatively small impact on our results ($\sim$20\% when tidal stripping is not considered and smaller than 10\% for the realistic case when tidal stripping is taken into account, see below). Therefore, for the sake of comparison with previous works, we do not include this dependence here and postpone its discussion to future work. By assuming that the subhalo mass function does not change within the halo, we can write the boost factor as
\begin{eqnarray}
B(M) & = & \frac{3}{{\cal L}_{\rm smooth}(M)} \, \int_{M_{\rm min}}^{M} \, \frac{dN(m)}{dm} \, dm \int_{0}^{1} dx_{\rm sub} \nonumber \\ 
& & \left[1+B(m)\right]  \, {\cal L}(m,x_{\rm sub}) \,  x_{\rm sub}^{2} ~ , 
\label{eq:BM}
\end{eqnarray}
where $dN(m)/dm$ is the subhalo mass function for a halo of mass $M$, $dN(m)/dm=A/M \left(m/M\right)^{-\alpha}$. The normalization factor is equal to $A=0.012$ for a slope of the subhalo mass function $\alpha=2$ and to $A=0.03$ for $\alpha=1.9$ \citep{Sanchez-Conde:2013yxa}, and was chosen so that the mass in the resolved substructure amounts to about 10\% of the total mass of the halo,\footnote{Extrapolating the subhalo mass function down to $m/M=10^{-18}$, those normalizations correspond to $\sim 50\%$ ($\sim 30\%$) of the total mass of the halo for $\alpha=2$ ($\alpha=1.9$). } as found in recent simulations \citep{Diemand:2007qr, Springel:2008cc}. Note that, as done in most of previous works,\footnote{See, e.g., \citet{Pieri:2009je} for one of the few exceptions.} we have not subtracted the subhalo mass fraction from the smooth halo contribution, so in principle, this leads to a slight overestimate of the smooth halo luminosity, and hence, to a slight underestimate of the boost factor. This is expected to be a small correction, though, since it applies mainly to the outer regions of the halo where the subhalos represent a larger mass fraction and the smooth contribution is much smaller and subdominant with respect to the contribution from substructure \citep{PalomaresRuiz:2010pn, SanchezConde:2011ap}.

In the case of an NFW profile, as the one we are using, the luminosity from the smooth DM distribution of a  field halo can also be expressed in terms of the maximum circular velocity, $V_{\rm max}^h$, \citep{Diemand:2008in}
\begin{equation}
{\cal L}_{\rm smooth}(V_{\rm max}^h)  \simeq \left(\frac{2.163}{f(2.163)}\right)^2 \frac{2.163 \, H_0}{12 \pi  G^2} \sqrt{\frac{c^h_{\rm V}(V_{\rm max}^h)}{2}} (V_{\rm max}^h)^3  ~,
\label{eq:LV}
\end{equation}
and, in a similar way, by including the radial dependence of the concentration of subhalos, one can obtain the subhalo luminosity function, ${\cal L}(V_{\rm max},x_{\rm sub})$.

In this case, the boost factor for a field halo with maximum circular velocity $V_{\rm max}^h$ (analogously to Eq.~(\ref{eq:BM})), can be written as
\begin{eqnarray}
B(V_{\rm max}^h)  & =  & \frac{3}{{\cal L}_{\rm smooth}(V_{\rm max}^h)} \, \int_{(V_{\rm max})_{\rm min}}^{V_{\rm max}^h} \, \frac{dN(V_{\rm max})}{dV_{\rm max}} \, dV_{\rm max}  \nonumber \\ 
 & & \int_{0}^{1} dx_{\rm sub} \, \left[1+B(V_{\rm max})\right] \, {\cal L}(V_{\rm max},x_{\rm sub}) \,  x_{\rm sub}^{2} ~, \nonumber \\
\label{eq:BV}
\end{eqnarray}
where $(V_{\rm max})_{\rm min}$ is the value of $V_{\rm max}$ which corresponds to $M_{\rm min}$.
In order to compute the luminosity in terms of $V_{\rm max}^h$ we need the subhalo mass function in terms of $V_{\rm max}$, and we use the result of \citet{Diemand:2008in}, $dN(V_{\rm max})/dV_{\rm max} = (0.108/V_{\rm max}^h) \, (V_{\rm max}^h/V_{\rm max})^4$.

\begin{figure}
	\hspace{-0.7cm}
	\includegraphics[width=270pt]{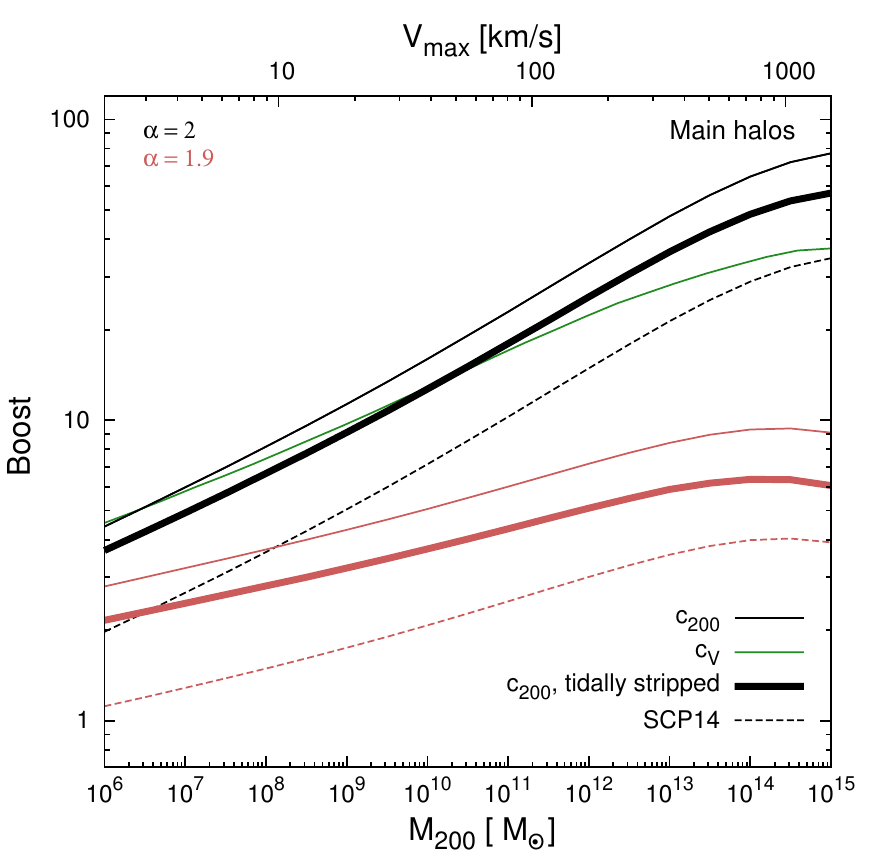}
	\caption{Halo substructure boost to the DM annihilation signal as a function of the host halo mass. We have used our $c_{200}(m_{200},x_{\rm sub})$ parametrization in Eq.~(\ref{eq:c200-fit}) and adopted $M_{\rm min}=10^{-6}\, M_{\odot}$. We present results for two values of the slope of the subhalo mass function, $\alpha=1.9$ (lower, light red lines) and $\alpha=2$ (black lines). We also show the boost obtained with the DM profile-independent definition of $c_{\rm V}$ (green line), for which we have used our fit for $c_{\rm V}(V_{\rm max},x_{\rm sub})$ in Eq.~(\ref{eq:cv-fit}), and $(V_{\rm max})_{\rm min} = 10^{-3.5}~{\rm km/s}$. Notably, the $c_{\rm V}$ result lies within the results found for $c_{200}$ and the two slopes of the subhalo mass function considered. Thin lines correspond to results obtained assuming subhalos and sub-subhalos are not truncated by tidal forces, while thick lines represent the more realistic case, in which subhalos and sub-subhalos have been tidally-stripped (see text). The dashed lines correspond to the results obtained in \citet{Sanchez-Conde:2013yxa} when assuming that both halos and subhalos of the same mass have the same concentration values.} 
	\label{fig:boost-csub}
\end{figure}

The results for the boost factor defined in Eqs.~(\ref{eq:BM}) and~(\ref{eq:BV}) are shown in Fig.~\ref{fig:boost-csub}, where we use the parametrizations for $c_{200} (m_{200},x_{\rm sub})$, $c_{\rm V} (V_{\rm max},x_{\rm sub})$, $c_{\rm V}^h (V^h_{\rm max})$ and $c^h_{200}(M_{200})$ given by Eqs.~(\ref{eq:c200-fit}), (\ref{eq:cv-fit})~(\ref{eq:cv-cal}) and P12, respectively. We depict the boost factor for field halos as a function of the halo mass and adopt $M_{\rm min}=10^{-6}\, M_{\odot}$ or, equivalently for an NFW profile, $(V_{\rm max})_{\rm min} = 10^{-3.5}$~km/s (thin solid lines). We show the results for both $c_{\rm V}$ (green line) and $c_{200}$ (in this case, for two values of the slope of the subhalo mass function, $\alpha=2$ and $\alpha=1.9$ with black and red lines, respectively). Both results are in good agreement, with the boost factor obtained from $c_{\rm V}$ lying within the boost factors obtained from $c_{200}$ for the two different slopes of the subhalo mass functions considered. The results obtained in \citet{Sanchez-Conde:2013yxa} are also shown (dashed lines). As done in this latter work and discussed above, we are including only the two first levels of substructure, namely subhalos and sub-subhalos, as the contribution of the third substructure level was found to be always less than $6\%$. Yet, we note that the second level (namely $B(m, x_{\rm sub})$ in our notation) can contribute up to $\sim40\%$ in some cases. As can be seen from Fig.~\ref{fig:boost-csub}, we obtain a total boost which is a factor of $2-3$ larger than that obtained in \citet{Sanchez-Conde:2013yxa}, where, we recall, the authors assumed that halos, subhalos and sub-subhalos of the same mass have the same concentrations. Interestingly, our results also agree well with those recently found by \citet{Bartels:2015uba} by means of a semi-analytical model for the boost based on mass-accretion histories and subhalo accretion rates. Similar boost values have also been reported in \citet{Zavala:2015ura}, where authors invoked the universality of DM clustering in phase space within subhalos across a wide range of host halo masses \citep{Zavala:2013lia} to predict DM annihilation signals.

We caution that, in our work and in \citet{Sanchez-Conde:2013yxa}, an NFW DM density profile is always assumed for all virialized structures. Nevertheless, it has been recently shown that subhalos and, very especially, microhalos with masses close to $M_{\rm min}=10^{-6}\, M_{\odot}$ seem to exhibit DM density profiles which are cuspier than NFW in the innermost regions \citep{Diemand:2008in, Ishiyama:2014uoa}. Thus, their concentrations do not correspond to the NFW concentration values discussed and adopted throughout this paper. Fortunately, it is possible to convert from one to another \citep{Ricotti:2002qu, Anderhalden:2013wd} and to perform a one-to-one comparison among them. The result of adopting subhalo concentrations that are corrected by the mentioned effect is a moderate increase of the boost factor, up to $\sim 30\%$ \citep{Anderhalden:2013wd, Ishiyama:2014uoa}.

\subsection{Effect of tidal stripping on the boost}  
\label{sec:boosttidal}

\begin{figure}
	\hspace{-0.7cm}
	\includegraphics[width=270pt]{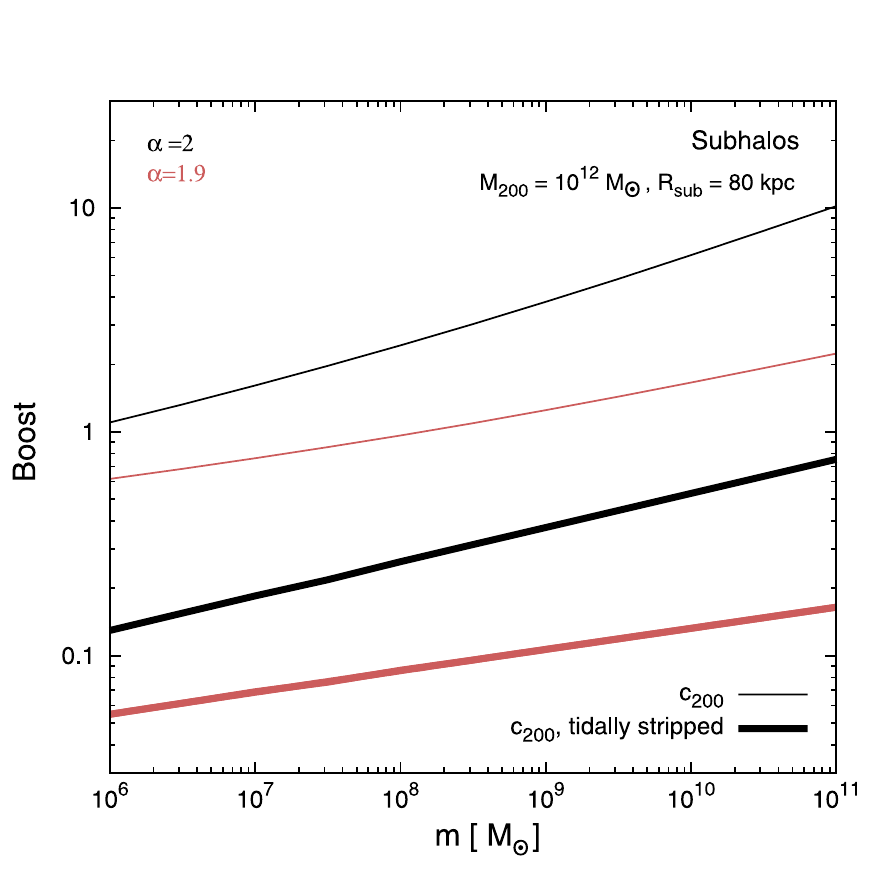}
	\caption{Example of subhalo substructure boost to the DM annihilation signal (the one expected, e.g., for dwarf satellite galaxies) as a function of the subhalo mass for the particular case of subhalos inside a host halo with mass $M_{200} = 10^{12} \, M_\odot$ and located at a distance of 80~kpc from the host halo center. This is approximately the case of Draco, one of the Milky Way dwarf galaxy satellites ($m_{\rm Draco} \sim \times 10^{8} \, M_\odot$ \citep{Lokas:2004sw}). We show results obtained assuming subhalos and sub-subhalos are not truncated (or, in some cases, destroyed) by tidal forces (thin lines), and assuming subhalos and sub-subhalos are tidally stripped (more realistic case; thick lines). We have used our $c_{200}(m_{200},x_{\rm sub})$ parametrization of Eq.~(\ref{eq:c200-fit}) and adopted $M_{\rm min}=10^{-6}\, M_{\odot}$. We also present results for two values of the slope of the subhalo mass function, $\alpha=1.9$ (light red lines) and $\alpha=2$ (black lines). See text for further discussion.} 
	\label{fig:subboost-csub} 
\end{figure}

So far in the calculation of the boost factor, we have not considered the fact that subhalos suffer from tidal forces within their host halos and thus, that they are expected to be truncated at some radius $r_t < r_{200}$. As already discussed above, this also implies that $m_{200}$ is not the true mass of the subhalo (which was nevertheless assumed to be such in the calculation of the boost factor in Sec.~\ref{sec:boostcomput}, Eqs.~(\ref{eq:BM}) and (\ref{eq:BV})). Therefore, a more precise value of the boost can be derived if the actual subhalo mass $m$, obtained by integrating the subhalo density distribution up to $r_t$, was adopted instead. In a similar way, the subhalo luminosity must be truncated at $r_t$ instead of $r_{200}$, i.e.,
\begin{eqnarray}
{\cal L}^{\rm t}_{\rm smooth}(m_{200},x_{\rm sub}) & \equiv & \int_0^{r_t} \rho^2_{\rm sub}(r) \, 4 \, \pi \, r^2 \, dr = \nonumber \\
& & \frac{m_{200} \,c_{200}^{3}(m_{200},x_{\rm sub})}{\left[f(c_{200}(m_{200},x_{\rm sub}))\right]^{2}} \frac{200 \, \rho_c}{9} \times \nonumber \\ 
& & \left(1-\frac{1}{(1+r_t/r_s(m_{200},x_{\rm sub}))^3} \right) ~.  \nonumber \\
\label{eq:LMt}
\end{eqnarray}
This is the only modification one has to include in the calculation of the boost up to the first level of substructures. However, to compute the boost factor of \emph{subhalos} (i.e., up to the second level of halo substructure), in addition to introducing the analogous modification in the calculation of the sub-subhalo luminosity, the variable $x_{\rm sub-sub} \equiv r_{\rm sub}/r_{200}$ (the equivalent to $x_{\rm sub}$ for sub-subhalos) must be substituted by $r_{\rm sub}/r_t$, where $r_{\rm sub}$ is the distance of the sub-subhalos to the center of the host subhalo. Moreover, we assume that tidal forces do not modify the subhalo and sub-subhalo mass functions per unit volume. This means that the number of sub-subhalos is reduced and therefore, the boost for subhalos.

A good approximation for the tidal radius is the so-called King radius \citep{King:1962wi}, defined as\footnote{Only strictly valid for point mass potentials for both the host halo and the subhalo, and for circular orbits.}
\begin{equation}
r_t = R_{\rm sub} \, \left( \frac{m_t}{3\,M(<R_{\rm sub})} \right)^{1/3} ~,
\label{eq:rtidal}
\end{equation}  
where $m_t$ is the true mass of the truncated subhalo (with radius $r_t$) located at a distance $R_{\rm sub}$ from the host halo center and $M(<R_{\rm sub})$ is the enclosed mass of the host halo out to that distance (and analogously for sub-subhalos).  {We have checked that this mass, $m_t$, does not largely differ on average from the actual subhalo mass found in simulations within statistical errors (see also, e.g., Fig.~15 of \citet{Springel:2008cc}).\footnote{ {We also note that different subhalo finders can lead to significantly different subhalo masses, see, e.g., \citet{Bosch:2014qsa}.}} We refer the interested reader to Appendix \ref{app:mtidal} for further details.}

The effect of tidal mass losses on the boost factor for halos is also shown in Fig.~\ref{fig:boost-csub}. As the subhalo luminosities are mainly dominated by the annihilations occurring in the innermost regions of subhalos, and these central regions are not significantly altered by tidal forces \citep{Kazantzidis:2003hb, Diemand:2007qr}, the impact of tidal stripping on the boost factor is expected to be moderate. Indeed, we find a suppression of about $20\%-30\%$, which mainly affects the second level of iteration. As mentioned above, tidal interactions do reduce significantly the number of sub-subhalos which are left inside a subhalo, thus, reducing the individual \emph{subhalo} boost (i.e., the one that would be applicable to, e.g., dwarf galaxies satellites of the Milky Way). 

We provide an example of the latter effect in Fig.~\ref{fig:subboost-csub}, which shows the subhalo boost factor as a function of subhalo mass calculated with (thick lines) and without (thin lines) tidal stripping, for the particular case of subhalos inside a host halo with mass $M = 10^{12} \, M_\odot$, and located at a distance of $R_{\rm sub} = 80$~kpc ($x_{\rm sub} = 0.39$) from the host halo center. This approximately corresponds to the case of the Draco dwarf spheroidal galaxy, which has a mass of the order of $10^{8} \, M_\odot$ \citep{Lokas:2004sw}. For this Galactocentric distance, and with very little dependence on the subhalo mass, the suppression of the subhalo boost factor is very significant, about 95\%, with respect to the case when tidal forces are not taken into account. Thus, we conclude that for the vast majority of dwarf galaxies in the Milky Way, the effect of substructure on the expected annihilation signal is probably at the level of a few tens of percent in the most optimistic cases, i.e., the largest subhalos and $\alpha=2$. We note that this result is also in line with that obtained by following the semi-analytical approach of \citet{Bartels:2015uba} for these same objects.

Finally, following \citet{Sanchez-Conde:2013yxa}, we also provide a parametrization of the boost factor values found for field halos (cf. Fig.~\ref{fig:boost-csub}). We do so only for the most realistic scenario in which subhalos (and sub-subhalos) are affected by tidal stripping as described in this section. We use the concentration-mass relation, $c_{200}(m_{200},x_{\rm sub})$, found in Sec.~\ref{sec:Nbody} and given by Eq.~(\ref{eq:c200-fit}). Our boost factor fits are valid in the halo mass range $10^{-6} < M_{200} \, [M_{\odot}] < 10^{15}$ and their accuracy is better than 5\% at all masses:
\begin{equation}
{\rm log}\, B(M) = \sum_{i=0}^{5}b_{i}\, \left[ {\rm log}\left( \frac{M}{M_{\odot}} \right) \right]^{i} ~.
\label{eq:boost-fit}
\end{equation} 
We provide the values of the fitted parameters, $b_i$, in Tab.~\ref{tab:boost-fit} for the two slopes of the subhalo mass function considered in our work, $\alpha=1.9$ and $\alpha=2$.

\begin{table} 
	\begin{center}
		\begin{tabular}{| c c c | }
			\hline
			$\alpha$ & 2 & 1.9  \\ \hline \hline
			$b_0$    & -0.186 & $-6.8\times 10^{-2}$ \\ \hline 
			$b_1$    &  0.144 & $9.4\times 10^{-2}$  \\ \hline  
			$b_2$    &  $-8.8 \times 10^{-3}$ & $9.8 \times 10^{-3}$  \\ \hline  
			$b_3$    &  $1.13 \times 10^{-3}$ & $1.05 \times 10^{-3}$  \\ \hline  
			$b_4$    &  $-3.7 \times 10^{-5}$ & $-3.4 \times 10^{-5}$  \\ \hline  
			$b_5$    &  $-2 \times 10^{-7}$ & $-2 \times 10^{-7}$ \\ \hline
		\end{tabular}
	\end{center}	
	\caption{Best-fit values of the parameters given in Eq.~(\ref{eq:boost-fit}) for the fit to the boost factor for field halos. We implicitly used the concentration-mass relation $c_{200}(m_{200},x_{\rm sub})$ of Eq.~(\ref{eq:c200-fit}) and included the effect of tidal stripping on both subhalos and sub-subhalos), as described in the text. These fits reproduce the thick lines in Fig.~\ref{fig:boost-csub}.}
	\label{tab:boost-fit}
\end{table}

\section{Summary} 
\label{sec:summary}

The internal structure of CDM halos, codified in the halo concentration parameter, has been extensively studied in the past and, indeed, multiple parametric forms have been proposed for the halo concentration-mass relation over a broad range of halo masses, for different cosmologies, etc. Yet, until now, no functional form had been proposed for the concentration-mass relation of subhalos, in part due to inherent difficulties, e.g., in properly defining and measuring subhalo concentrations and masses. In this work, we made use of public data from two Milky Way-size N-body cosmological simulations with a superb subhalo statistics, namely VL-II \citep{Diemand:2008in} and ELVIS \citep{Garrison-Kimmel:2013eoa}, to overcome some of such difficulties (Sec.~\ref{sec:Nbody}). All together, these simulations made possible to characterize the structural properties of subhalos with masses $10^6 - 10^{11}~h^{-1} M_{\odot}$. We first described the subhalo concentration $c_{\rm V}$ based on a profile-independent definition in terms of $V_{\rm max}$ and $R_{\rm max}$, and showed how this is related to the usual $c_{200} (m_{200})$ (Sec.~\ref{sec:definitions}). Then, we investigated in detail how the concentration of subhalos varies as a function of mass, maximum circular velocity and distance to the host halo center, which allowed us to provide accurate fits for the subhalo concentration (the first ones to our knowledge) encapsulating the mentioned dependences (Sec.~\ref{sec:fits}). Qualitatively, and as already shown in previous works, we found the subhalo concentration i) to slowly decrease with increasing mass and ii) to significantly increase towards the host halo center for subhalos of the same mass. Since VL-II and ELVIS do not provide data at the smallest and largest scales, we also made use of the microhalo simulations by \citet{Ishiyama:2014uoa} and the BolshoiP large-scale structure simulation \citep{Klypin:2014kpa, Rodriguez-Puebla:2016ofw}, respectively, to approximately account for the expected field halo and subhalo behavior at the two extremes of the full CDM (sub)halo hierarchy. Furthermore, we studied the scatter of the concentration of subhalos in both VL-II and ELVIS in different mass and radial bins (Sec.~\ref{sec:scatter}). Interestingly, we found the subhalo-to-subhalo scatter to be similar to that obtained for field halos, i.e., of order $0.10-0.12$ dex.

Our subhalo concentration results are particularly relevant for computing the boost to the DM annihilation signal due to the presence of halo substructure (Sec.~\ref{sec:boost}). Since the substructure boost is very sensitive to the way the structure of subhalos is modeled at all masses (or equivalently, at all maximal circular velocities), our detailed characterization of subhalo concentrations is expected to have an important impact on indirect DM searches. This is particularly true now that the field has reached maturity,\footnote{In the form of robust constraints on the DM annihilation cross section that rule out, for the first time using gamma rays, the so-called \emph{thermal} value for light DM masses \citep{Ackermann:2015zua}. Note, though, that this thermal value had already been reached before by making use of measurements of the cosmic microwave background anisotropy (for recent analyses, see, e.g., \citet{Lopez-Honorez:2013lcm, Diamanti:2013bia, Madhavacheril:2013cna, Ade:2015xua, Slatyer:2015jla, Kawasaki:2015peu}).} and is therefore calling for more accurate predictions of the expected DM annihilation flux from astrophysical targets (such as dwarf galaxies or galaxy clusters). In previous works, as a first order approximation, the concentration of subhalos was considered to be equal to that of field halos of the same mass. However, it was already known that subhalos exhibit larger concentrations~\citep{Ghigna:1999sn, Bullock:1999he, Diemand:2007qr, Diemand:2008in} and this, in turn, implies a larger substructure boost factor. Indeed, using our fits for the subhalo concentration-mass relation, we found boosts that are typically a factor of $2-3$ larger than previous results (see Sec.~\ref{sec:boostcomput} and, particularly, Fig.~\ref{fig:boost-csub}), as those given by the model of \citet{Sanchez-Conde:2013yxa}. Similar results have also been recently found in \citet{Bartels:2015uba} and \citet{Zavala:2015ura} by following different but complementary approaches. In particular, \citet{Bartels:2015uba} adopted a semi-analytical approach which includes information about both halo and subhalo mass-accretion histories, subhalo accretion times, and consistently accounts for tidal stripping in the calculation of the boost, while \citet{Zavala:2015ura} used a novel statistical measure of DM clustering in phase space within subhalos to predict the boosts.

This is not the end of the story though. Unavoidable tidal forces affecting the subhalo population remove mass from the outer parts of subhalos and, as a result, they are truncated when compared to field halos of the same mass~\citep{Ghigna:1998vn, Kazantzidis:2003hb, Diemand:2006ik, Diemand:2007qr}. When taking tidal stripping into account (Sec.~\ref{sec:boosttidal}), the final boost factor for \emph{field} halos (such as our own Milky Way) is reduced by a factor of about $\sim 30\%$. Yet, as also noticed by \cite{Bartels:2015uba}, this suppression is much more significant when considering the boost factor for \emph{subhalos} (like those hosting the dwarf satellite galaxies of the Milky Way). In that case, the removal of sub-substructure from subhalos results in total boosts that are only at the level of a few tens of percent in the most optimistic cases (we illustrated this in Fig.~\ref{fig:subboost-csub} for the case of subhalos located at a distance of 80~kpc from the center of their $10^{12} \, M_\odot$ host halo, similar to the case of Draco in the Milky Way). Finally, we provided a parametrization of the boost factor for field halos that can be safely applied over a wide range of halo masses, and that should be considered as a refinement of the model of \citet{Sanchez-Conde:2013yxa}. 

Although this work is a necessary step towards a reliable and definitive substructure boost model, some important issues still remain for future work. In particular, it would be desirable to build the model directly from simulation data at \emph{all} subhalo mass scales. Complementing the VL-II and ELVIS results obtained here with other existent or upcoming high-resolution N-body simulations should make possible this further refinement of the model. The precise slope of the subhalo mass function (and its possible dependences with, e.g., host halo mass \citep{Jiang:2014nsa, Hellwing:2015upa}) is another source of uncertainty that should be explored in detail in the future. Finally, we note that ours and previous substructure boost models rely on results from pure DM-only simulations. It is unclear at the moment the role that baryons might have in this context (see, e.g., the recent works by \citet{Fiacconi:2016cih, Wetzel:2016wro}), and in particular their impact on the substructure boost values.

\section{Acknowledgments}
We are very grateful to J\"urg Diemand, Tomoaki Ishiyama and Anatoly Klypin for useful discussions during the completion of this work. We also thank Shea Garrison-Kimmel for his help with the ELVIS simulations. AM has been supported by the Funda\c{c}\~ao para a Ci\^encia e a Tecnologia (FCT) of Portugal and thanks SLAC and IFIC for hospitality. AM has been also partially supported by the FCT through the project PTDC/FIS-NUC/0548/2012. MASC is a Wenner-Gren Fellow and acknowledges the support of the Wenner-Gren Foundations to develop his research. He also acknowledges the support from the MULTIDARK project of Spanish MCINN Consolider-Ingenio: CSD2009-00064. MASC is very grateful to SLAC and the Instituto de F\'isica Te\'orica in Madrid, where part of this work was done. SPR is supported by a Ram\'on y Cajal contract, by the Spanish MINECO under Grants No. FPA2014-54459-P and SEV-2014-0398 and by the Generalitat Valenciana under Grant No. PROMETEOII/2014/049. AM and SPR are also partially supported by the Portuguese FCT through the CFTP-FCT Unit 777 (PEst-OE/FIS/UI0777/2013). \\

\appendix

\section{Truncated subhalo radius and mass}  \label{app:mtidal}

In this appendix, we detail the procedure we followed to compute the truncated subhalo radius, $r_t$, given by Eq.~(\ref{eq:rtidal}) in the main text and also reproduced here for convenience:

\begin{equation}
r_t = R_{\rm sub} \, \left( \frac{m_t}{3\,M(<R_{\rm sub})} \right)^{1/3} ~,
\label{eq:apprtidal}
\end{equation}  

For each subhalo of mass $m_{200}$ located at a distance $R_{\rm sub}$ from the host halo center, we calculate its concentration $c_{200}(m_{200},x_{\rm sub})$ by making use of our fit in Eq.~(\ref{eq:c200-fit}), where $x_{\rm sub}=R_{\rm sub}/R_{200}$. Next, we compute the host halo mass within $x_{\rm sub}$, i.e., $M(<R_{\rm sub}$) assuming the DM density distribution of the host halo to follow an NFW profile. Finally, we solve Eq.~(\ref{eq:apprtidal}) to obtain $r_t$ and $m_{200}(<r_t) \equiv m_t$. 

\begin{figure}
	\hspace{-0.7cm}
	\includegraphics[width=0.50\textwidth]{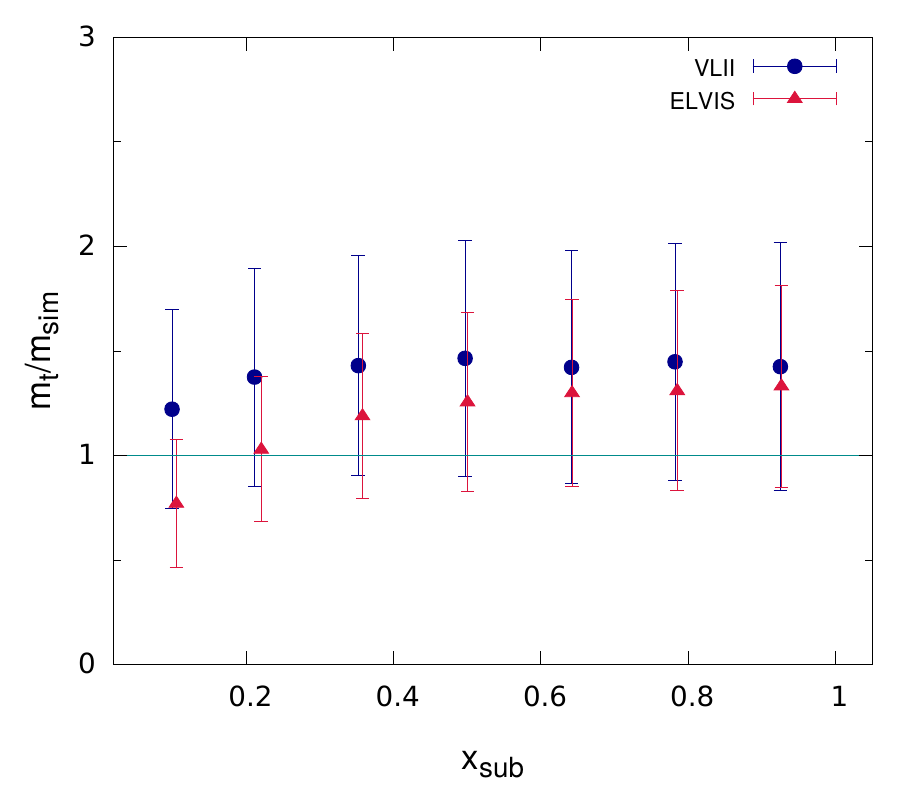}\\
	\includegraphics[width=0.49\textwidth]{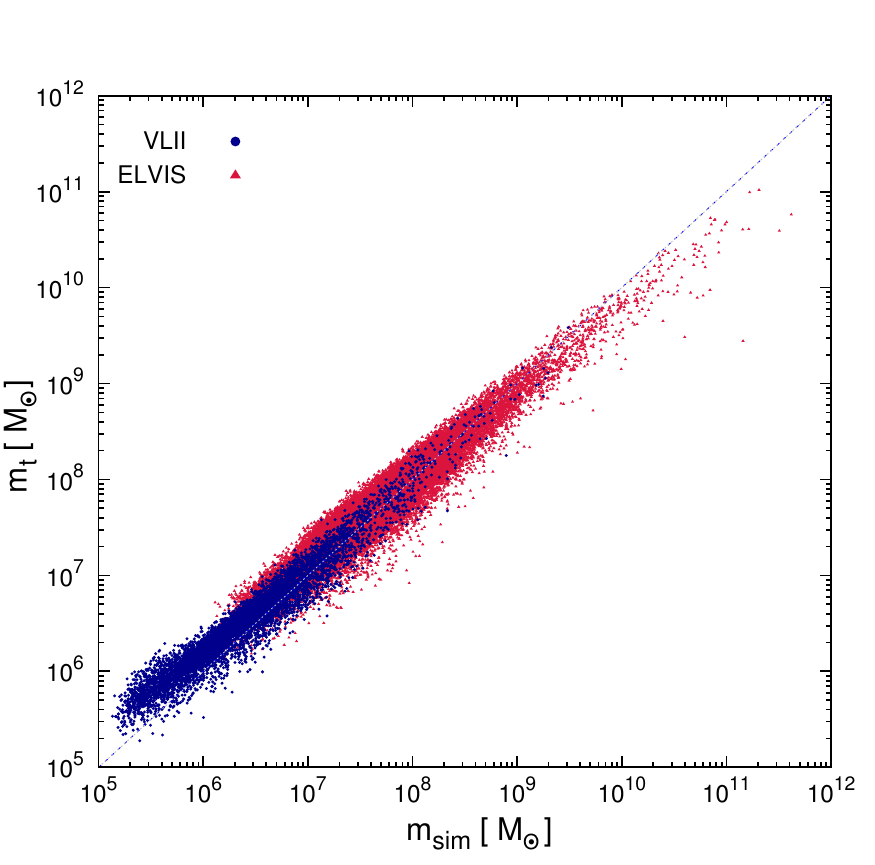}
	\caption{Actual subhalo masses, $m_{\rm sim}$, as found in the VL-II and ELVIS simulations (blue and red symbols, respectively) compared to the truncated subhalo masses, $m_t$,  computed according to Eq.~(\ref{eq:apprtidal}). {\it Top:} dependence of $m_t/m_{\rm sim}$ on subhalo distance to host halo center, with corresponding $1\sigma$ error bars. {\it Bottom:} $m_t$ versus $m_{\rm sim}$ with no information on subhalo location inside the host. } 
	\label{fig:Msubcomparison} 
\end{figure}

It is important to note that this calculation of $r_t$ implicitly relies on the assumption that subhalo DM density profiles can be well described by NFW up to $r_t$, as this is the main assumption behind Eq.~(\ref{eq:c200-fit}) for the subhalo mass-concentration relation. Although valid as a fair first order approximation, we once again recall that the actual subhalo DM density profile may differ substantially from NFW (even within $r_t$). Thus, it would be interesting to compare the truncated subhalo masses computed in the way described above, $m_t$, to the actual subhalo masses found in the simulations, $m_{\rm sim}$. We do so in Fig.~\ref{fig:Msubcomparison} for both ELVIS and VL-II. 

The top panel of Fig.~\ref{fig:Msubcomparison} depicts the $m_t/m_{\rm sim}$ ratio and shows that, with our method to estimate subhalo masses, we typically overestimate the actual mass by $\sim30-40\%$ at all distances from the center of the host halo, except in the innermost regions for the case of ELVIS. It is also noticeable the large spread of $m_t/m_{\rm sim}$ values, with the associated $1\sigma$ error bars always encompassing mass ratios equal to one, and reaching values up to a factor two in VL-II.

The bottom panel of Fig.~\ref{fig:Msubcomparison} shows a direct comparison of $m_t$ versus $m_{\rm sim}$ instead, that does not take into account the position of the subhalo inside the host. From this figure we seem to underestimate (overestimate) the actual subhalo masses at the largest (smallest) subhalo masses, differences being of the order of those shown in the top panel.

\bibliography{bibliosubhalos_arXiv}

\end{document}